\documentclass[aps,twocolumn,prb,amsmath,amssymb,showpacs,floatfix,altaffilletter,superscriptaddress,longbibliography]{revtex4-2}
\usepackage{graphicx}
\usepackage{times}
\usepackage[varg]{txfonts}
\usepackage{bm}
\usepackage{amsmath}
\usepackage{sansmath}
\usepackage{xspace}
\usepackage{xr}
\usepackage{natbib} 
\usepackage{xcolor}
\usepackage{braket}
\usepackage{subfigure}

\usepackage{changes}

\usepackage[percent]{overpic}

\usepackage[colorlinks,bookmarks=false,citecolor=blue,linkcolor=red,urlcolor=blue]{hyperref}

\newcommand{\beq}{\begin{equation}}
\newcommand{\eeq}{\end{equation}}

\newcommand{\ie}[0]{i.e.}
\newcommand{\eg}[0]{e.g.}

\begin{document}

\title{Dissipation-induced long-range order in the one-dimensional Bose-Hubbard model}

\author{Afonso L. S. Ribeiro}
\affiliation{CeFEMA-LaPMET, Departamento de F\'{ı}sica, Instituto Superior T\'{e}cnico, Universidade de Lisboa, Av.~Rovisco Pais, 1049-001 Lisboa, Portugal}

\author{Paul McClarty}
\affiliation{Max Planck Institute for the Physics of Complex Systems, N\"{o}thnitzer Str. 38, 01187 Dresden, Germany}

\author{Pedro Ribeiro}
\affiliation{CeFEMA-LaPMET, Departamento de F\'{ı}sica, Instituto Superior T\'{e}cnico, Universidade de Lisboa, Av.~Rovisco Pais, 1049-001 Lisboa, Portugal}
\affiliation{Beijing Computational Science Research Center, Beijing 100193, China}

\author{Manuel Weber}
\affiliation{Institut f\"{u}r Theoretische Physik and W\"{u}rzburg-Dresden Cluster of Excellence ct.qmat, Technische Universit\"{a}t Dresden, 01062 Dresden, Germany}


\begin{abstract}
Understanding the stability of strongly correlated phases of matter when coupled to environmental degrees of freedom is crucial for identifying the conditions under which these states may be observed. Here, we focus on the paradigmatic one-dimensional Bose-Hubbard model, and study the stability of the Luttinger liquid and Mott insulating phases in the presence of local particle exchange with site-independent baths of non-interacting bosons. We perform a numerically exact analysis of this model by adapting the recently developed wormhole quantum Monte Carlo method for retarded interactions to a continuous-time formulation with worm updates; we show how the wormhole updates can be easily implemented in this scheme. For an Ohmic bath, our numerical findings confirm the scaling prediction that the Luttinger-liquid phase becomes unstable at infinitesimal bath coupling. We show that the ensuing phase is a long-range ordered superfluid with spontaneously-broken U(1) symmetry. While the Mott insulator remains a distinct phase for small bath coupling, it exhibits diverging compressibility and non-integer local boson occupation in the thermodynamic limit. Upon increasing the bath coupling, this phase undergoes a transition to a long-range ordered superfluid. Finally, we discuss the effects of super-Ohmic dissipation on the Luttinger-liquid phase. Our results are compatible with a stable dissipative Luttinger-liquid phase that transitions to a long-range ordered superfluid at a finite system-bath coupling.
\end{abstract}

\maketitle

\section{Introduction} 

It is a truism that isolated quantum systems do not exist in nature except perhaps in a cosmological setting. The system generally consists of those degrees of freedom that are monitored or controlled while the environment is, in principle, everything else. Exactly what constitutes the system and the environment is highly dependent on context. For a single atom the environment might include the radiation field. When the system is a many-particle system, such as a lattice of interacting localized moments in a solid, the environment could include any nuclear spin bath, phonons or itinerant electrons as well as the surrounding contacts, cryostat and so on. 

Coupling between the system and the environment tends to result in entanglement build-up between them if there was none before. Frequently physicists have the luxury of overlooking the complexities of the dynamical processes involved as their net effect is to cause the system to thermalize and one may then study the system at equilibrium as being at some finite temperature set by the (comparatively large scale) surroundings. Indeed it is understood, in a many-particle context, that thermalization generically occurs in subsystems of isolated quantum systems \cite{srednicki1994chaos,deutsch2008,rigol2008thermalization,alessio2016}. The rather typical emergence of statistical mechanics is, of course, a remarkable feature of natural laws but there are instances where reducing a complex environment to a mere thermostat misses other crucial qualitative effects of the environmental degrees of freedom on the system. 

Over the last several decades there have been considerable efforts to increase the characteristic time scales associated to system-environment coupling precisely in order to study nearly isolated quantum systems. In the early days these efforts concentrated on few-body quantum systems such as atoms in cavities in which decoherence and quantum dissipative processes were witnessed in detail for the first time \cite{raimond2001}. On the theoretical side this involved the study of by now classic models such as Caldeira-Leggett and spin-boson models \cite{PhysRevLett.46.211,Leggett1987}. In recent years these efforts have focussed more on quantum simulation and computation where the desired system has a large number of degrees of freedom. The trend therefore is towards greater control of larger and larger quantum systems and greater influence over the nature of the environment.

While technological developments make these issues more pressing, the broad question of how a quantum bath can affect a many-body system is a long-standing one. Beyond the issues of thermalization the bath is known to be able to influence the system in various other ways. For example, a phonon bath can induce a spin-Peierls transition in a spin chain where the chain translational symmetry breaks down \cite{Bray1983}. More recently, significant efforts have gone into studying the effects of dissipation on Luttinger liquids \cite{CastroNeto1997,Cazalilla2006,Lobos2012,Cai2014,PhysRevB.97.035148,friedman2019dissipative,Weber2022,Danu2022,PhysRevB.107.165113,majumdar2023localization,Kuklov2023,weber2023quantum,Martin2023}. Broadly speaking the Luttinger liquid can be destabilized by coupling to a bath leading either to long-range order, disordered phases that are suggestive of a novel dissipation induced phase, or even phases with long-range order with no obvious analogue in closed systems.
Indeed, dissipation can induce nontrivial intermediate-coupling fixed points that may undergo fixed-point annihilation. This physics is known to appear in single spins coupled to a bath \cite{PhysRevLett.108.160401,PhysRevB.106.L081109,PhysRevLett.130.186701,hu2022kondo}, but has also been predicted for extended spin chains \cite{Martin2023}.
 
Quantitative results have been obtained for different dissipation mechanisms using quantum Monte Carlo (QMC) simulations: particle dissipation was studied for a chain of interacting hard core bosons where each site was coupled to an independent bath described by a one-dimensional system itself \cite{PhysRevB.97.035148}; while in this study the Luttinger liquid phase seemed unaffected, the coupling to the gapless bath turned the Mott phase into a compressible insulator.
Other QMC approaches considered the coupling to a bath of harmonic oscillators that, when integrated out, leads to a retarded long-range interaction in the system degrees of freedom.
Away from half filling, energy dissipation via a retarded density-density interaction can induce a gapless liquid phase with vanishing superfluid density that is therefore distinct from the Luttinger liquid obtained at zero dissipation \cite{Cai2014}.
In the one-dimensional spin one-half Heisenberg chain, Ohmic dissipation, which couples isotropically to all spin components, induces antiferromagnetic long-range order \cite{Weber2022} by suppressing the fluctuations that would usually destroy it in one dimension \cite{Mermin1966,Hohenberg1967}. 
Dissipation-induced order has
also been studied in chains of coupled quantum rotors using classical Monte Carlo \cite{2005JPSJ...74S..67W,PhysRevB.83.115134,PhysRevB.85.214302}.
The broader message of these studies is that the bath can nontrivially affect a quantum system in a way that departs from simple thermalization. When the system is gapless the stability of any phase in the isolated system should not be taken for granted once coupled to a gapless bath. Also, when the system is gapped, coupling to the bath can induce perturbative departures from the behavior of the isolated system of a qualitative nature.

\begin{figure}[t]
    \centering
    \begin{overpic}[width=0.9\linewidth]{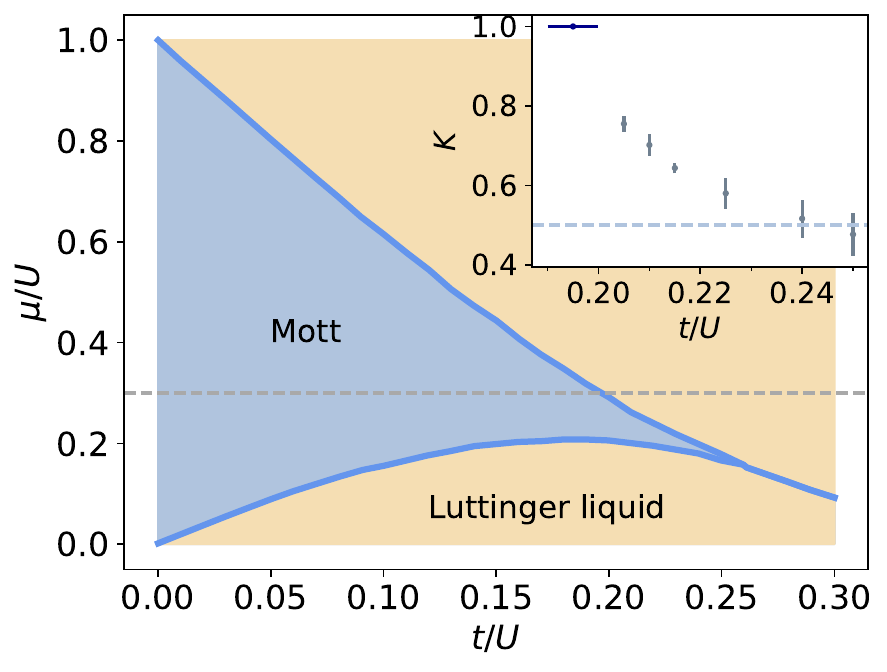}
    \end{overpic}
    \\ \hspace{0.1cm}
    \begin{overpic}[width=0.95\linewidth]{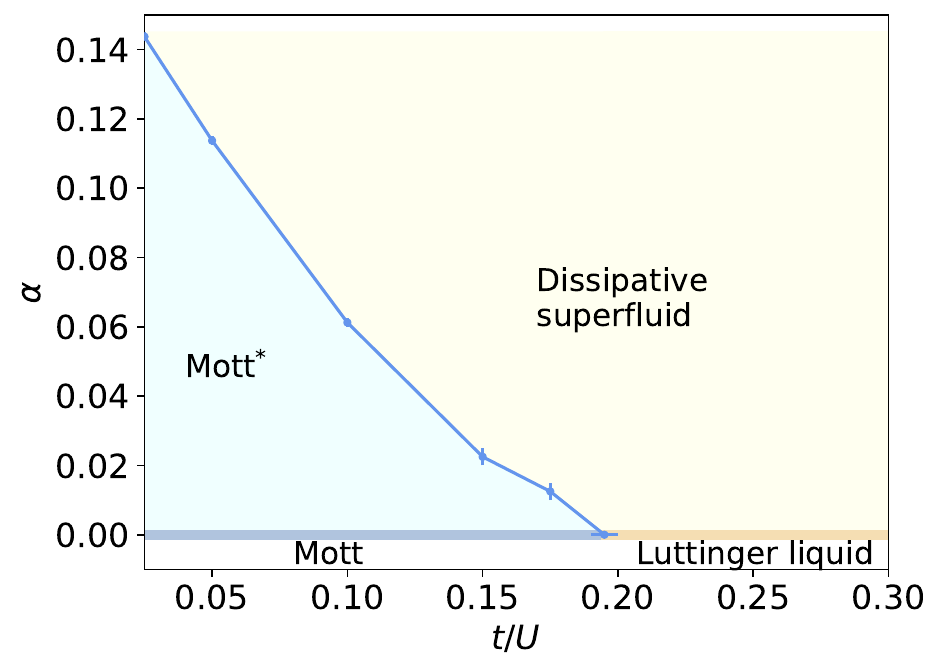}
    \put(85,135.5){\large(a)}
    \put(85,63){\large(b)}
    \end{overpic}
    \caption{(a) Phase diagram of the Bose-Hubbard chain, as obtained from the density-matrix renormalization group (DMRG) in Ref.~\cite{kiely2022superfluidity}, as a function of hopping $t$ and chemical potential $\mu$. We focus on the first Mott lobe with fixed onsite boson number $n_i=1$ and the transition towards the quasi-superfluid Luttinger-liquid phase.
    The dashed line depicts $\mu / U = 0.3$, along which we perform our QMC simulations.
     Inset: Luttinger parameter extracted from QMC for $\mu/U=0.3$ with dashed line at $K=1/2$ which is marginal for bath exponent $s=1.75$. The $K=1$
point indicates the uncertainty in the numerically established phase boundary.    (b) Numerically established phase diagram for $\mu/U=0.3$ in the presence of an Ohmic bath ($s=1$). At finite dissipation strengths $\alpha$, we find a dissipative superfluid with long-range order and an insulating but infinitely-compressible state which we call Mott$^\ast$ that is perturbatively connected to the Mott insulator.}
    \label{fig:bhphases}
\end{figure}

In this paper, we explore the effects of an independent oscillator bath coupled to each site of a one-dimensional Bose-Hubbard model via single-particle exchange. The Bose-Hubbard model is one of the canonical models of condensed matter physics \cite{giamarchi2003quantum,cazalilla2011one}. In one dimension, the zero temperature phase diagram of this model has a quasi-superfluid or Luttinger liquid phase when the kinetic energy to interaction strength is large. In the opposite limit, there is a series of Mott insulating lobes as a function of the chemical potential. The phase diagram for the isolated chain around the first Mott lobe is illustrated in Fig.~\ref{fig:bhphases}(a). The model is also of experimental relevance. In one dimensional arrays of Josephson junctions, in the quantum limit, composed of superconducting islands separated by insulating junctions there is an effective hopping between neighboring islands and an interaction term whose strength and range depends on the junction capacitance. When the average boson number is high this model maps to the Bose-Hubbard model and the arrays themselves exhibit an analogue of a metal-insulator transition with a coherent phase on one side and a phase with localized charges on the other \cite{Zwerger2003,Bruder2005}. The dissipation in these systems depends on the resistance of the junctions \cite{Zwerger1989}. A second realization of the Bose-Hubbard model is in optical traps of bosonic atoms \cite{Jaksch1998}. In such systems, the atoms are subject to a periodic potential and a trapping potential whose depth introduces a competition between coherent hopping and repulsion within each potential well.  A Mott to superfluid transition has been observed experimentally in three dimensional traps \cite{Greiner2002}.
In principle, dissipation similar in nature to that we are considering here could be achieved by coupling each site of a Bose-Hubbard quantum simulator to an independent chain of non-interacting bosons \cite{PhysRevB.97.035148}.

The Bose-Hubbard model and the bath coupling are introduced in detail in Section~\ref{Sec:BHandbath}. In order to study the open Bose-Hubbard chain we have developed a continuous-time path-integral QMC method with worm updates \cite{Prokofev:1998aa}. In Section~\ref{sec:QMC}, we demonstrate how the recently-developed wormhole moves \cite{PhysRevB.105.165129} for retarded dissipative interactions can be implemented easily in this scheme. Our results are described and discussed in Section~\ref{sec:results}. We show that the Luttinger liquid is unstable to infinitesimal bath coupling, consistent with a power counting argument, that furthermore leads to a long-range ordered superfluid phase. The Mott phase gives way to an infinitely-compressible phase, denoted Mott$^*$, that undergoes a transition for larger bath coupling into the superfluid. Our findings for the Ohmic bath are summarized in the phase diagram in Fig.~\ref{fig:bhphases}(b). Moreover, we will discuss the effects of super-Ohmic dissipation on the Luttinger liquid, for which power counting predicts stable and unstable regimes to dissipation. Finally, in Sec.~\ref{Sec:Conclusions} we summarize our results and conclude.

\section{The Bose-Hubbard Model and the Bath\label{Sec:BHandbath}}

\subsection{Bose-Hubbard chain\label{Sec:BHmodel}}

The Bose-Hubbard model in one dimension is one of the classic models of quantum many-body physics whose properties have been summarized in detail in the literature \cite{giamarchi2003quantum,cazalilla2011one}. Its Hamiltonian includes nearest-neighbor hopping, onsite repulsion, and a chemical potential:
\beq
\label{eq:BH}
H_s = 
-t \sum_{i} \left( b^\dagger_i b_{i+1} + {\rm h.c.} \right) 
+ \frac{U}{2} \sum_{i} b^\dagger_i b_i^\dagger b_i b_i 
- \mu  \sum_{i} b^\dagger_i b_i
\, .
\eeq
Here, the single-particle operators $b^\dagger_i$ ($b_i$) create (annihilate) a boson at lattice site $i\in\{1, \dots, L\}$.
The phase diagram of this model is naturally parametrized by $\mu/U$ and $t/U$. Considering first the $t/U=0$ limit, the remaining terms mutually commute and minimizing the onsite energy $\epsilon_i =  \left(U/2\right) n_i\left(n_i-1\right) - \mu \, n_i  $ reveals that the occupation number $n_i = \langle b_i^\dagger b_i \rangle$ changes discontinuously from $n_i-1$ to $n_i$ when $\mu/U = n_i-1$. There is also a gap to changing the local boson number, \ie, the compressibility defined as $\kappa = \partial n_i/\partial \mu =0$. Taken together these facts point to this being a Mott phase. The gap implies that this phase is stable to small hopping $t$. The steps in average occupation at zero hopping result in a series of Mott lobes as $\mu$ is varied. The shape of the first Mott lobe ($n_i=1$) in the $t$-$\mu$ plane is illustrated in Fig.~\ref{fig:bhphases}(a).

 At the upper (lower) border of the Mott phase  as $\mu$ is varied it becomes favorable to add (subtract) particles and the density changes continuously and the compressibility becomes nonzero. In contrast, at the tip of each Mott lobe increasing $t/U$ it becomes favorable to overcome the energy gap by allowing bosons to hop through the lattice. It turns out that the transition is a Berezinskii-Kosterlitz-Thouless transition at the tip and a mean-field transition elsewhere.

The line of transitions out of the Mott phase connects to a Luttinger liquid with linearly dispersing excitations and a static correlator that falls off as a power law, \ie,
\beq
\label{eq:spcor}
\langle b_i^\dagger b_{j} \rangle \sim \vert i-j \vert^{-K/2} \, ,
\eeq
where $K$ is the Luttinger parameter that goes to zero in the $t/U\rightarrow \infty$ limit. Within the Luttinger liquid the parameter $K$ approaches universal values at the transition lines depending on the Mott lobe and whether the line is at the top, bottom or tip of the lobe.

\subsection{Coupling to the bath\label{Sec:Bath}}

We now supplement the system Hamiltonian $H_s$ with a coupling to a bath, \ie, $H=H_s + H_b + H_{sb}$.
At each site $i$ of the chain we introduce an independent ensemble of bosonic oscillators with creation (annihilation) operators $a_{iq}^\dagger$  ($a_{iq}$)  where $q$ runs over a continuum of modes with frequency $\omega_q$. The oscillators are taken to be mutually non-interacting:
\beq
H_b = \sum_{iq} \omega_{q} \, a^\dagger_{iq}  a_{iq}
\, .
\eeq 
In this work, we consider the coupling between the Bose-Hubbard chain and the bath to be
\beq
\label{Eq:Hsb}
H_{sb} = \sum_{iq} \lambda_q \left( b^\dagger_{i} a_{iq} + {\rm h.c.} \right)
\eeq
chosen to preserve the symmetries of the chain including translation symmmetry and the internal U(1) symmetry associated to boson number conservation of system plus bath. Our numerical method is tailored to this choice of single-particle dissipation. Another natural system-bath coupling takes the form of a coupling to the bath density, $\sum_{iq} b^\dagger_i b_i \, (a^\dagger_{iq}+a_{iq})$, which has been studied for a system of hard core bosons \cite{Cai2014}. 
The bath spectral function $J(\omega)= \pi \sum_q \lambda^2_q \, \delta(\omega -\omega_q)$ is chosen to be of power-law form,
\beq
J(\omega) = 2\pi \alpha \omega_{c}^{1-s} \omega^s
\, ,
\qquad
0 < \omega \leq \omega_\mathrm{c} \, ,
\label{eq:spectraldensity}
\eeq
where $\alpha$ is a dimensionless coupling and $\omega_c$ a cutoff frequency, beyond which $J(\omega)$ is zero. An Ohmic bath corresponds to exponent $s=1$ and a super-Ohmic bath has $s>1$.

As the bath is non-interacting with a linear coupling to the system, it may be integrated out leaving an effective retarded interaction between the system bosons of the form
\beq
\label{Eq:Sret}
\mathcal{S}_{\rm ret} = - \iint_0^\beta d\tau d\tau' \, b^\dagger_i(\tau) \, D(\tau-\tau') \, b_{i}(\tau')
\eeq
that is mediated by the bath propagator
\beq
D(\tau) = \int_0^{\omega_c} d\omega \frac{J(\omega)}{\pi}
\frac{e^{-\omega \tau}}{1-e^{-\beta \omega}} 
\, ,
\qquad
0 \leq \tau < \beta
\, .
\eeq
Here, $D(\tau+\beta)=D(\tau)$ and $\beta=1/T$ is the inverse temperature. For the power-law spectrum in Eq.~\eqref{eq:spectraldensity},
$D(\tau) \sim 1/\tau^{1+s}$ for $\omega_c \tau \gg 1$.

\section{Quantum Monte Carlo method}
\label{sec:QMC}

For our simulations, we developed a continuous-time QMC method with wormhole updates, which have been introduced recently in the context of the directed-loop algorithm \cite{PhysRevB.105.165129}. In the following, we will show how the wormhole updates can be implemented in the worm algorithm \cite{Prokofev:1998aa}, which has advantages in the presence of strong diagonal interactions, as is the case for the Bose-Hubbard model for large $U/t$.

\subsection{Interaction representation and worm algorithm}

In the interaction picture, we split the Hamiltonian $H = H_0 - V$ into an unperturbed part $H_0$ that is diagonal in the real-space occupation-number basis and an off-diagonal perturbation $V$, such that the Dyson expansion of the partition function becomes
\beq
\label{Eq:Dyson}
Z  = \mathrm{Tr} \, e^{-\beta H_0}\sum_{m=0}^\infty \,
\int_0^\beta d\tau_{m} \dots \int_0^{\tau_2} d\tau_1 \,
V(\tau_m) \dots V(\tau_1)
\, .
\eeq
Here, we have defined $V(\tau) = e^{\tau H_0} V e^{-\tau H_0}$.
Because the operator sequence $V(\tau_m) \dots V(\tau_1)$ is time-ordered for a given expansion order $m$, it can be interpreted as performing an imaginary-time evolution of an initial state $\ket{\alpha_0}$, for which each operator $V(\tau_\ell)$ propagates the state from $\ket{\alpha_{\ell-1}}$ to $\ket{\alpha_\ell}$. The states $\ket{\alpha_\ell}$ are many-body eigenstates of $H_0$ in the occupation-number basis with eigenenergies $E^0_\ell$ so that only the application of $V$ can change the state. We assume that the operators $V$ are non-branching, \ie, each initial state is mapped to exactly one final state. Then, the trace over all states $\ket{\{\alpha\}}$ reduces to a sum over the initial state $\ket{\alpha_0}$, as the trace requires $\ket{\alpha_m} = \ket{\alpha_0}$. As a result, the partition function has become a sum over world-line configurations that describe the propagation of a set of particles in imaginary time.

Path-integral QMC methods provide an efficient way to sample the sum over all world-line configurations.
A Monte Carlo configuration is described by the set $\mathcal{C}=\{\alpha_0, m, \mathcal{C}_m\}$ for which all vertex labels are contained in $\mathcal{C}_m = \{\nu_1, \dots, \nu_m\}$.
Eventually, the partition function becomes
\beq
\label{Eq:MCexpansion}
Z  
= \sum_{\alpha_0} e^{-\beta E^0_0}\sum_{m=0}^\infty \sum_{\nu_1 \dots \nu_m}
\prod_{\ell = 1}^{m} W_{\nu_\ell}
= \sum_{\mathcal{C}}
W(\mathcal{C})
\, .
\eeq
In particular, $W(\mathcal{C})$ factorizes into a product of vertex weights $W_{\nu_\ell}=
\braket{\alpha_{\ell}|V_{\nu_\ell}|\alpha_{\ell-1}}$.
Each vertex is characterized by a set of variables $\nu=\{\Gamma, \tilde{\nu}\}$, where $\Gamma$ distinguishes different types of vertices and $\tilde{\nu}$ contains the internal variables of each vertex. For example, for the isolated Bose-Hubbard model in Eq.~\eqref{eq:BH}, $V$ is given by the hopping term, such that we
set the label $\Gamma = \textnormal{"hop"}$ and have $\tilde{\nu}=\{i, \tau\}$. The vertex weights become
\beq
W^{\mathrm{hop}}_{\tilde{\nu}_{\ell}} = t \,
e^{\tau_{\ell} \Delta E^0_\ell}
\braket{\alpha_{\ell}| \big(b^\dagger_{i_\ell} b_{i_\ell+1} + {\rm h.c.}\big) |\alpha_{\ell-1}}
\, ,
\eeq
where $\Delta E^0_\ell = E^0_{\ell}-E^0_{\ell-1}$.
The vertex weights are strictly positive and the matrix elements are non-branching. The diagonal energies of the Bose-Hubbard model have been defined in Sec.~\ref{Sec:BHmodel}. The kinetic term of the Bose-Hubbard model leads to a single hopping event between neighboring sites, which is also referred to as a kink in the world-line configuration.

To efficiently update the world-line configurations, we use the worm algorithm \cite{Prokofev:1998aa}.
It operates in an extended configuration space
\beq
Z_\mathrm{ext} = Z + C_W \sum_{i_A i_B} \int d\tau_A \int d\tau_B \, G_{i_A i_B}(\tau_A, \tau_B)
\eeq
which is supplemented by the single-particle Green's function
\begin{align}
\nonumber
G_{i_A i_B}(\tau_A, \tau_B) =
\frac{1}{Z} &
 \mathrm{Tr} \, e^{-\beta H_0}\sum_{m=0}^\infty \frac{1}{m!}
\int_0^\beta d\tau_m \dots \int_0^{\beta} d\tau_1 \\ & \times \mathcal{T}_\tau \, V(\tau_m) \dots V(\tau_1) \,
b_{i_A}(\tau_A) \, b^\dagger_{i_B}(\tau_B) \, .
\end{align}
Here, $\mathcal{T}_\tau$ is the time-ordering operator.
It is apparent that this interaction expansion still has a world-line representation, but the two single-particle operators lead to world-line discontinuities with additional weights
\beq
W^{-}_A = e^{\tau_A \Delta E^0_{A}} \braket{\alpha'_{A} | b_{i_A}|\alpha_{A}} \, ,
\quad
W^{+}_B = e^{\tau_B \Delta E^0_{B}} \braket{\alpha'_{B} | b^\dagger_{i_B}|\alpha_{B}} \, ,
\eeq
which have to be incorporated appropriately in the product of all weights. In the following, we will briefly summarize the ideas of the worm algorithm, but omit to describe the algorithmic details, which can be found in the standard literature on this method \cite{Prokofev:1998aa, POLLET20072249, 10.21468/SciPostPhysCodeb.9}. Our Julia implementation closely follows the open-source code described in Ref.~\cite{10.21468/SciPostPhysCodeb.9}.

We now summarize the basic updates using the notation of Ref.~\cite{10.21468/SciPostPhysCodeb.9}.
To switch from the partition function to the Green's function sector, we use an update called INSERTWORM, which inserts a pair of creation and annihilation operators right after each other at a random position. We choose one of the two operators as the worm head which can now perform a random walk through space and time.
To this end, we can apply the MOVEWORM update, which moves the worm by some distance $\Delta \tau$ that is determined by the diagonal energies in the exponential factors. Moreover, the worm has the option to insert a new kink and hop to its neighboring site via the INSERTKINK update, but also to delete an existing kink via a reverse process described by the DELETEKINK update. For the Bose-Hubbard model with local occupation numbers larger than one, a situation can occur where a kink cannot be deleted; then we use the PASSINTERACTION update, in which the worm head moves straight through the kink.
Once the worm head returns to its tail, we can use the GLUEWORM update to remove the world-line discontinuities and return to the partition function sector. Each of the individual updates are local moves that scale as $\mathcal{O}(\beta^0 L^0)$. As the average expansion order $\langle m \rangle \propto L \beta$, the worm algorithm scales linearly in the system parameters in order to generate statistically independent Monte Carlo configurations.
For further details, see Ref.~\cite{10.21468/SciPostPhysCodeb.9}.

\subsection{Wormhole updates for retarded interactions}

In the following, we want to discuss how the coupling to a bath that exchanges particles with the system, as defined in Sec.~\ref{Sec:Bath}, can be simulated with the worm algorithm. If the system only couples to one bath mode, Eq.~\eqref{Eq:Hsb} is just another hopping term that can be treated as described before. If the bath modes approach a continuum, it is not a priori clear how to choose the weights correctly to hop into the right mode. One way to simulate an Ohmic bath is to couple each site to an independent one-dimensional chain \cite{PhysRevB.97.035148} which then has to be simulated together with the system. A convenient way to simulate quantum dissipative interactions is to use retarded interactions, as the one in Eq.~\eqref{Eq:Sret}. So far, the worm algorithm could only deal with diagonal retarded interactions \cite{Cai2014}, but recently the novel wormhole updates \cite{PhysRevB.105.165129} have been introduced in the framework of the directed-loop algorithm which can also simulate off-diagonal retardation efficiently. Here we show how to implement the wormhole updates within the worm algorithm.

To treat the spin-boson interaction in the interaction picture, we include $H_b$ in the unperturbed part $H_0$ and $H_{sb}$ in the perturbation $V$. Then, the Dyson expansion can be performed as in Eq.~\eqref{Eq:Dyson} and we obtain separate traces over the system and bath degrees of freedom. As has been described in detail in Ref.~\cite{PhysRevB.105.165129}, the quadratic bath modes can be traced out exactly using Wick's theorem. After reorganizing the perturbation expansion, this results in an effective retarded interaction of the form given in Eq.~\eqref{Eq:Sret}.
Alternatively, one can use the coherent-state path integral to integrate out the bath, perform the Dyson expansion, and map everything back to an operator formalism \cite{PhysRevLett.119.097401}.

We include the retarded interaction vertex \eqref{Eq:Sret} in the perturbation expansion of Eq.~\eqref{Eq:MCexpansion} by defining a new vertex with label $\Gamma = \textnormal{"ret"}$. Its vertex variables are $\tilde{\nu} = \{i,\omega,\tau,\tau'\}$
and the vertex weight is given by
\begin{align}
\nonumber
W^\mathrm{ret}_{\tilde{\nu}_\ell} =
D(\omega_\ell,\tau_\ell - \tau'_{\ell}) \,
& e^{\tau_{\ell} \Delta E^0_\ell}
\braket{\alpha_{\ell}| b^\dagger_{i_\ell} |\alpha_{\ell-1}} \\
\times \, &e^{\tau'_{\ell} \Delta E'^0_{\ell}}
\braket{\alpha'_{\ell}| b_{i_{\ell}} |\alpha'_{\ell-1}} 
\, .
\end{align}
Here we use $\ell$ as a label for the full vertex and distinguish different stages of the propagated states via
$\ket{\alpha_\ell}$ and $\ket{\alpha'_\ell}$.
In addition to the two imaginary-time variables of the retarded interaction, we also want to sample the spectral function $J(\omega)$ and so the frequency $\omega$ is included in the vertex. To this end, we write the bath propagator
\beq
D(\omega,\Delta\tau)
    =
    \frac{2\alpha \omega_c}{s} \,
    \mathcal{J}(\omega) \, P(\omega,\Delta\tau)
\eeq
in terms of the probability distribution functions in $\omega$ and $\Delta\tau$,
\begin{gather}
\mathcal{J}(\omega) = \frac{J(\omega)/\omega}{\int d\omega J(\omega)/\omega}
= s \, \omega_c^{-s} \, \omega^{s-1} \, ,
\\
P(\omega,\Delta\tau) =
\frac{\omega \, e^{-\omega \Delta\tau}}{1-e^{-\beta \omega}} \, ,
\end{gather}
respectively \cite{PhysRevB.105.165129}. Both functions can be sampled using inverse transform sampling. Using uniformly-distributed random numbers
$\xi, \xi'\in[0,1)$,
we first draw $\omega = \omega_c \left(1-\xi\right)^{1/s}$ and then, using this frequency, sample the time difference $\Delta\tau = - \ln[1-\xi'(1-e^{-\beta\omega})]/\omega$,
as derived in Ref.~\cite{PhysRevB.105.165129}.
Note that the sampled time difference has a direction, because the retarded interaction in Eq.~\eqref{Eq:Sret} is not symmetric under $\tau \leftrightarrow \tau'$.

\begin{figure}[t]
    \centering
    \begin{overpic}[width=0.9\linewidth]{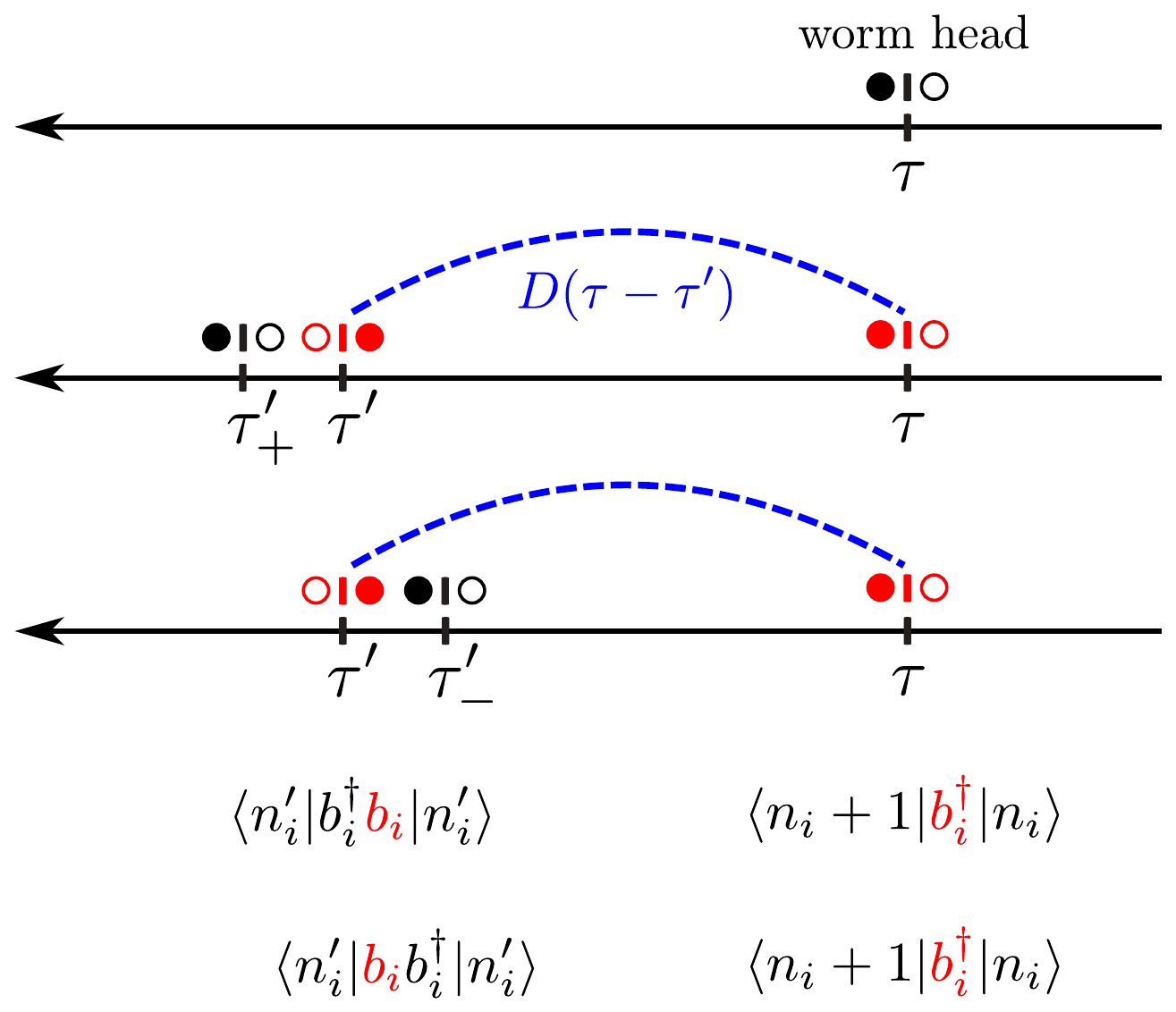}
    \put(1,81){\large(a)}
    \put(1,61){\large(b1)}
    \put(1,39.5){\large(b2)}
    \put(1,15.0){\large(c1)}
    \put(1,2.2){\large(c2)}
    \end{overpic}
    \caption{Schematic illustration of the INSERTWORMHOLE update.
    (a) We consider the boson creation operator $b_i^\dagger$ as the worm head (black) sitting at the space-time coordinate $(i,\tau)$.
    (b) To perform a nonlocal jump of the worm head, we formally transform it into one subvertex of the retarded interaction (red) and create a new world-line discontinuity at time $\tau'$. The annihilation operator (red) completes the retarded interaction vertex, and the worm head (black) is included with equal probability at time $\tau'_{\pm}$ right before or after the second subvertex, as illustrated by options (b1) and (b2). The time difference is chosen according to the bath propagator $D(\tau-\tau')$ (dashed blue line), which connects the two subvertices. (c) Expectation values of the operators contained in (b1) and (b2), for which state propagation is from right to left. For simplicity, we neglected additional factors to the weight like the time-dependent exponentials or the bath propagator. Note that the weight at time $\tau$ does not change during the wormhole update.
    }
    \label{fig:wormholes}
\end{figure}
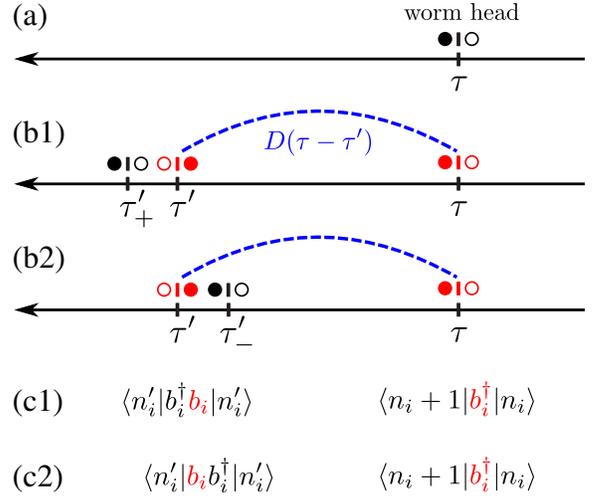

We use the Metropolis algorithm to formulate the novel INSERTWORMHOLE update, under which the worm head jumps nonlocally in imaginary time and thereby creates a retarded interaction vertex, as illustrated in Fig.~\ref{fig:wormholes}.  The acceptance probability
$
A(\mathcal{C} \to \mathcal{C}') = \min[1,R(\mathcal{C} \to \mathcal{C}')]
$
is determined by the ratio
\beq
R(\mathcal{C} \to \mathcal{C}') =
\frac{W(\mathcal{C}') \, T(\mathcal{C}'\to \mathcal{C})}{W(\mathcal{C}) \, T(\mathcal{C}\to \mathcal{C}')}
\eeq
of Monte Carlo weights $W(\mathcal{C})$ and $W(\mathcal{C}')$ before and after the update and the transition amplitudes $T$ between the two configurations. Let us assume that the worm head is a creation operator at site $i$ and time $\tau$ [Fig.~\ref{fig:wormholes}(a)]. After the wormhole update, this operator has turned into the first subvertex of the retarded interaction and we have to include an annihilation operator as the second subvertex at time $\tau'$. The time difference is obtained from sampling $\mathcal{J(\omega)}$ and $P(\omega,\tau-\tau')$ as described before. The worm head can be included at time $\tau'_{\pm}$ right before or after $\tau'$, which we choose with probability $1/2$ [Figs.~\ref{fig:wormholes}(b1) and \ref{fig:wormholes}(b2)]. As a result, the transition amplitude becomes
\beq
T(\mathcal{C}\to \mathcal{C}')
=
\frac{1}{2} \,\mathcal{J}(\omega) \, d\omega \,
P(\omega, \tau-\tau') \, d\tau' 
\, \delta(\tau' - \tau'_\pm)
\, .
\eeq
For the reverse process of removing a retarded interaction,
\beq
T(\mathcal{C}' \to \mathcal{C}) = 
\delta(\tau' - \tau'_\pm)
\, .
\eeq
The ratio of the Monte Carlo weights is mainly determined by the local weights that are affected by the wormhole update, \ie, the worm head and the included retarded interaction vertex. For the two possibilities of putting the worm head at $\tau'_\pm$, we combine the weights outlined in Figs.~\ref{fig:wormholes}(c1) and \ref{fig:wormholes}(c2), and obtain
\begin{align}
\nonumber
\frac{W(\mathcal{C}')}{W(\mathcal{C})}
&=
\frac{W^{+}_{i \tau'_{\pm}}
W^{\mathrm{ret}}_{i\omega \tau \tau'} \, d\omega \, d\tau \, d\tau'}
{W^{+}_{i \tau} \, d\tau} \\
&= \frac{\alpha \omega_c}{s} 
\braket{n'_i| (2 \, b^\dagger_{i} b_{i} + 1 \mp 1) |n'_i}
\nonumber\\
& \qquad\qquad \times  \mathcal{J}(\omega) \,P(\omega, \tau-\tau') \, d\omega \, d\tau' d\tau'_{\pm} \, .
\end{align}
Note that the time-dependent exponential factors drop out, because during this update time values do not get shifted and $\tau'_{\pm} \to \tau'$. Eventually, we obtain the acceptance ratio
\beq
R(\mathcal{C} \to \mathcal{C}') =
\frac{2\alpha \omega_c}{s}  \left(2n'_i +1 \mp 1 \right) \, .
\eeq
Because we have proposed the time difference according to the bath propagator, the time and frequency dependence of the retarded interaction do not affect the acceptance probability.
Note that the acceptance ratio only depends on the occupation $n'_i$ of the propagated state at time $\tau'$ before we include the new operator and the choice of inserting the worm head before or after the subvertex. The occupation at time $\tau$ has already been taken care of by the worm head before the wormhole update and therefore drops out. The inverse acceptance ratio $R(\mathcal{C}' \to \mathcal{C}) = 1/R(\mathcal{C} \to \mathcal{C}')$ applies to the REMOVEWORMHOLE update. If the worm head approaches a subvertex that is the same operator, the REMOVEWORMHOLE update cannot be applied, so that we choose the PASSWORMHOLE update to pass this operator with probability one. A similar analysis can be performed if we choose the boson annihilation operator as the worm head.

The wormhole updates can be easily included in an existing implementation of the worm algorithm. The biggest algorithmic change concerns the linking of the vertices. In the original formulation, only connections between the different kinks need to be considered while the worm head traverses the world-line configuration. In the presence of the wormholes, we also need to include nonlocal connections in time, which makes it necessary to find appropriate positions in the operator sequence. As a result, the formal computational cost of the wormholes will be slightly slower than linear in $\beta$.
The calculation of observables in the system's degrees of freedom is as usual, whereas bath observables are not accessible directly, but can be recovered from the distribution of retarded vertices using generating functionals \cite{PhysRevB.94.245138}.
For example, the average expansion order of retarded vertices,
$\langle m_\mathrm{ret} \rangle = - \beta \, \langle H_{sb} \rangle/2$, is directly related to the spin-boson interaction energy.

\section{Results}
\label{sec:results}

\subsection{Power Counting}
\label{ss:powercounting}

\begin{figure}[t]
    \centering
    \includegraphics[width=\linewidth]{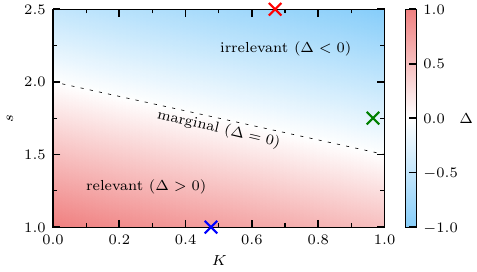}
    \caption{Stability of the Luttinger-liquid phase of the isolated system to an infinitesimal bath coupling according to power counting.
    As a function of the Luttinger parameter $K$ and the bath exponent $s$, we can distinguish two regimes where the bath is an (ir)relevant perturbation separated by a line where it is marginal. The color coding indicates the scaling dimension $\Delta$ given in Eq.~\eqref{eq:scalingdim} and the three markers the parameter sets studied throughout this paper.
    }
    \label{fig:powers}
\end{figure}

We begin our exploration of the phases of the dissipative Bose-Hubbard model by giving a preview coming from power counting arguments \cite{cardy_1996}. 
We consider the retarded interaction in Eq.~\eqref{Eq:Sret} as a weak perturbation to the isolated system.
For $\omega_c \tau \gg 1$, it decays like $D(\tau-\tau') \sim 1/ \vert \tau -\tau' \vert^{1+s}$. Moreover, the correlator $\langle b^\dagger(x,\tau) \, b(x,\tau') \rangle \sim\ |\tau - \tau'|^{-K/2}$ in the Luttinger liquid, as dictated by Eq.~\eqref{eq:spcor} and conformal invariance.
We now rescale space and time coordinates $x \rightarrow bx$ and $\tau \rightarrow b^z \, \tau $. 
As we are interested in instabilities out of the Luttinger liquid we take dynamical exponent $z=1$ noting that $z$ may depart from this value at transitions into the Mott phase.
Then, the scaling dimension of this perturbation to the Luttinger liquid, 
\beq
\label{eq:scalingdim}
\Delta = 2-s-\frac{K}{2} \, ,
\eeq
leads to the marginal bath exponent at $\Delta=0$,
\beq
s_\mathrm{marg} = \frac{1}{2} \left( 4 - K \right) \, ,
\eeq
so that, for 
$s > s_\mathrm{marg}$,
the bath is irrelevant ($\Delta < 0$) and the Luttinger liquid is stable to its presence and in the opposite limit
the bath is relevant ($\Delta > 0$) and the renormalization-group flow goes to a dissipation-induced fixed point.
An overview over the two regimes and the corresponding scaling dimensions is given in Fig.~\ref{fig:powers}.

We know that the Luttinger parameter varies with bare couplings $t/U$ and $\mu/U$ from $0$ to $1$ within the quasi-superfluid phase. In particular, along the Mott to quasi-superfluid phase boundary $K=1$ apart from the cusp where $K=1/2$. The $K=1/2$ contour bounds the phase boundary meeting it only at the cusps. The $K=1$ marginal bath exponent is $s=1.5$ and the $K=1/2$ marginal bath exponent is $s=1.75$, as visible in Fig.~\ref{fig:powers}. For an Ohmic bath with $s=1$, the bath is relevant everywhere in the quasi-superfluid phase which is therefore expected to be unstable to an infinitesimal coupling to this gapless bath. For $s>2$ the bath is always irrelevant leaving the Luttinger liquid stable at small couplings.
For $s=1.75$, we expect three phases for small bath coupling, as discussed in more detail below.
The parameter sets studied throughout this paper are indicated by the three markers in Fig.~\ref{fig:powers}.

\subsection{Closed system}

\begin{figure*}[t]
    \centering
    \begin{overpic}[width=0.45\linewidth]{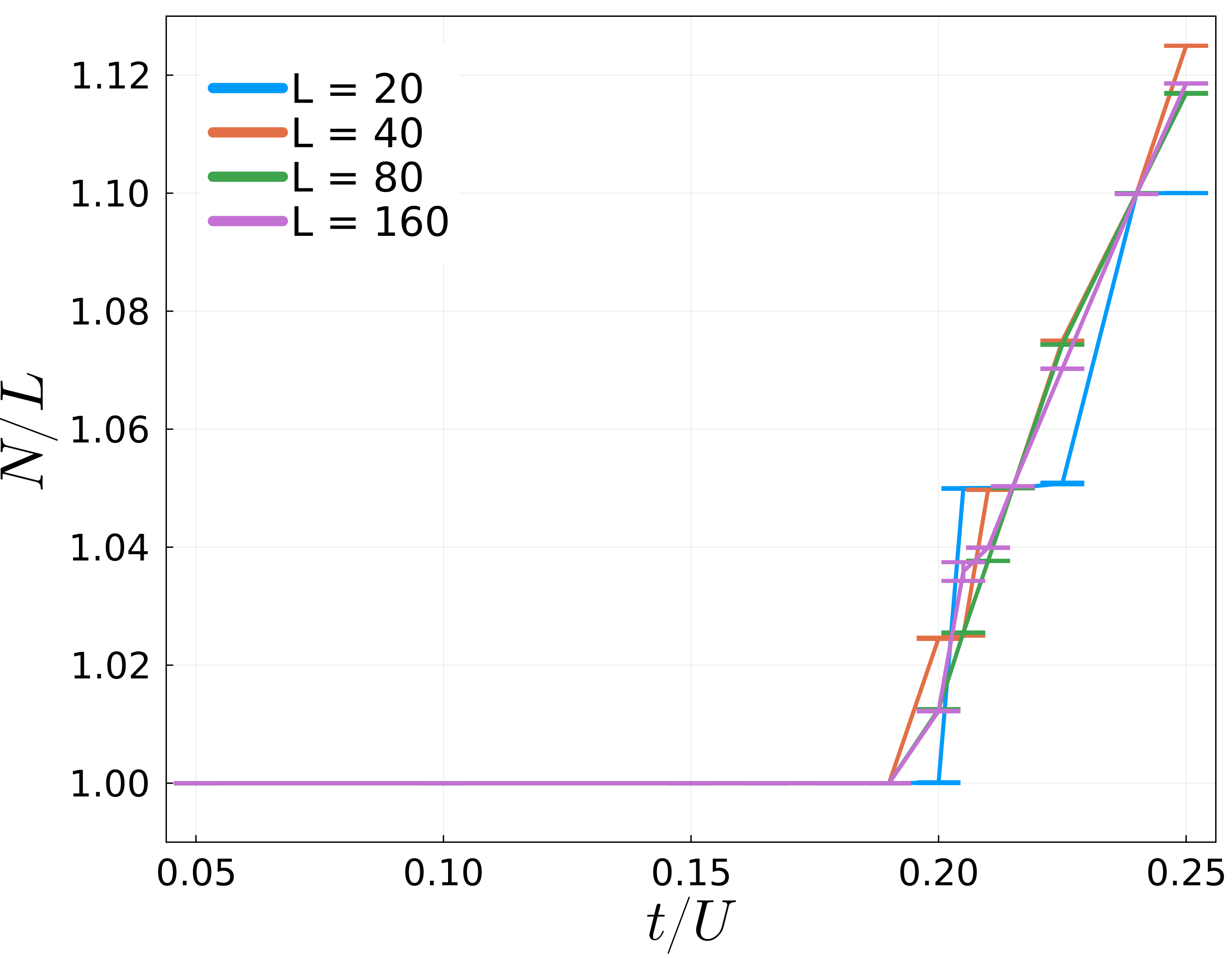}
    \put(88,15){\large(a)}
    \end{overpic}
    \hspace{0.5cm}
    \begin{overpic}[width=0.45\linewidth]{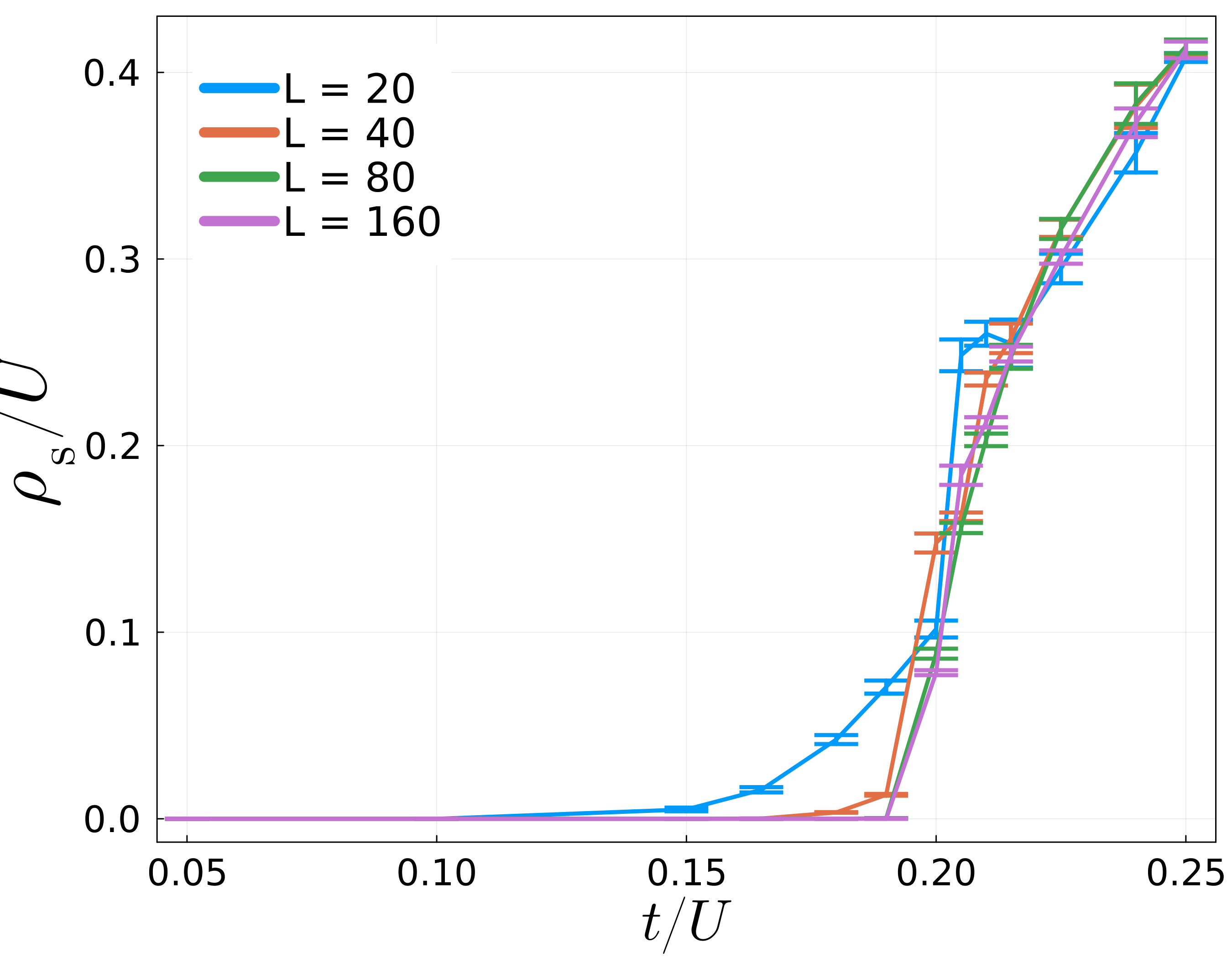}
    \put(88,15){\large(b)}
    \end{overpic}
    \begin{overpic}[width=0.45\linewidth]{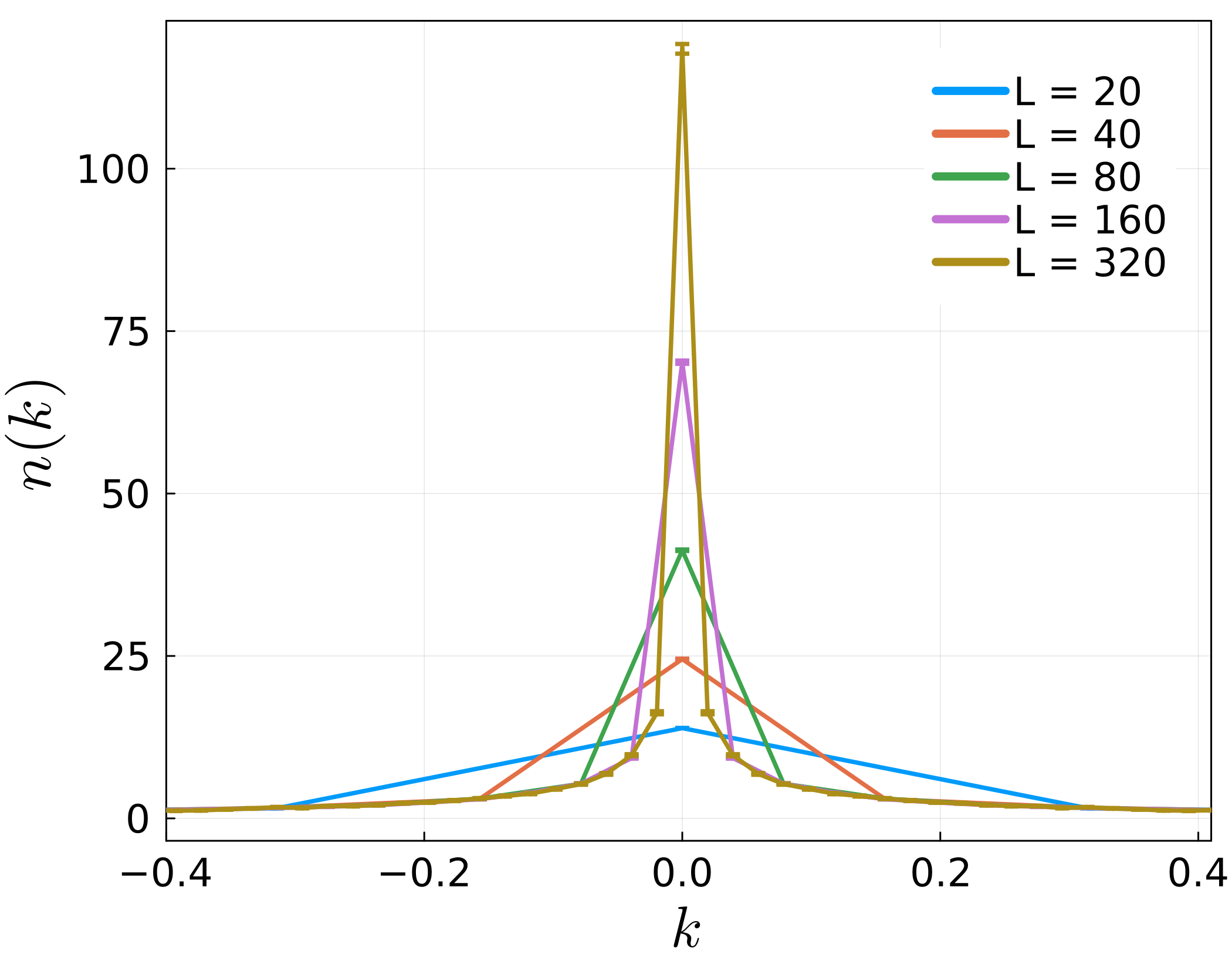}
    \put(88,15){\large(c)}
    \end{overpic}
    \hspace{0.5cm}
    \begin{overpic}[width=0.45\linewidth]{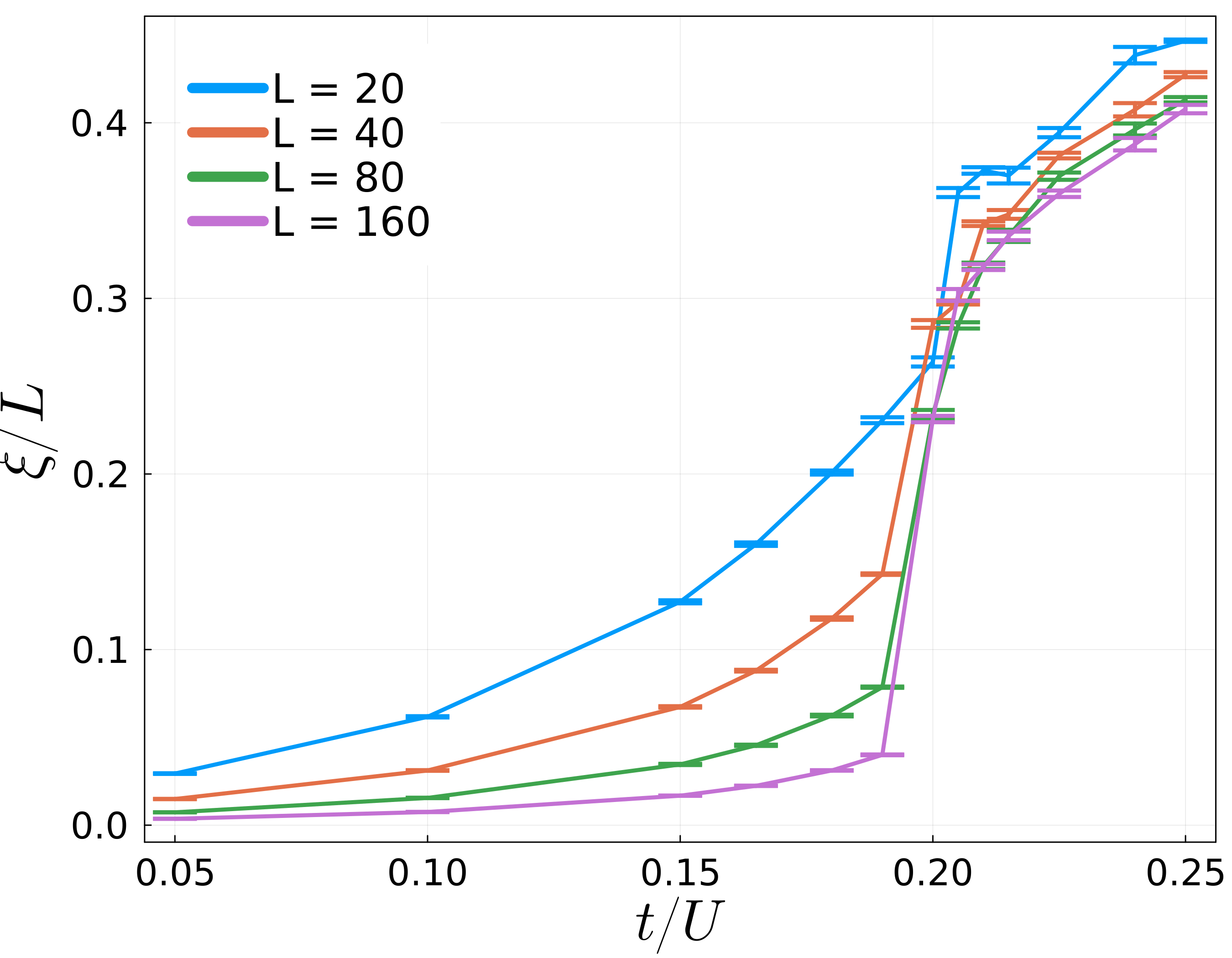}
    \put(88,15){\large(d)}
    \end{overpic}
    \caption{Overview over different observables for the isolated Bose-Hubbard chain.
    (a) Average particle number per site, $N/L$, and (b) superfluid stiffness $\rho_\mathrm{s}$ as a function of hopping $t/U$ and for different system sizes $L$. (c) Finite-size dependence of the momentum distribution function $n(k)$ at $t/U=0.25$. (d) Correlation length $\xi/L$ as a function of $t/U$. Here, $\mu/U=0.3$.}
    \label{fig:closed_overview}
\end{figure*}

To further set the scene for our numerical work, we remark on numerical results for zero bath coupling.  Figure~\ref{fig:bhphases}(a) shows the phase boundary of the first Mott lobe of the non-dissipative Bose-Hubbard model determined using DMRG and taken from Ref.~\cite{kiely2022superfluidity}. The QMC results reported here focus on the line $\mu/U=0.3$ (dashed line in the figure). Along this line, the boundary between the Mott phase and the quasi-superfluid is at $t_c/U\approx 0.195$ as determined using both DMRG~\cite{kiely2022superfluidity} and our QMC method.

We prepare for our analysis of the dissipative model by first recalculating some well-known properties of the isolated chain. When the dynamical critical exponent, $z$, is known, the exploration of low-temperature properties using QMC is carried out using finite-size scaling at fixed $\beta/L^z$. Here, coupling to a bath may lead $z$ to depart from its values at zero bath coupling. We therefore converge the results carefully at large $\beta$. To get an intuition about possible finite-size effects that occur at zero temperature, we also do this for the non-dissipative system. For all our simulations, we use periodic boundary conditions and restrict the boson occupation to $n_i \leq 10$, which is large enough that it does not affect our results.

The existence of the Mott phase can be determined from the particle density $N/L = \sum_i n_i/L$ which is strictly equal to one in the first Mott lobe and which varies continuously in the Luttinger-liquid phase. This is confirmed by our QMC simulations, as shown in Fig.~\ref{fig:closed_overview}(a). While $N/L$ is equal to one in the Mott phase without visible fluctuations, the particle density starts to increase beyond $t_c/U \approx 0.195$. For small system sizes, we observe that the particle density evolves in a series of step functions as we increase the hopping. Each of these steps corresponds to an integer filling of our lattice with bosons, for which particle fluctuations are frozen out at zero temperature. Because at finite system sizes, the ground states within neighboring particle-number sectors are nearly degenerate, we need to tune a system parameter (here the hopping) by a finite amount to switch into the next sector. For a detailed discussion of this finite-size effect, see Appendix~\ref{App:FSS_isolated}.

A direct probe for the Luttinger-liquid phase is the superfluid stiffness. In world-line QMC simulations, it can be accessed directly via the winding-number
fluctuations
\cite{PhysRevB.36.8343}, \ie,
\beq
\rho_\mathrm{s} = \frac{L}{\beta} \, \langle W^2 \rangle \, .
\eeq
Our QMC results in Fig.~\ref{fig:closed_overview}(b) confirm that $\rho_\mathrm{s}$ scales to zero in the Mott phase, but is finite in the Luttinger-liquid phase. Note that the finite-size effects from switching between the different particle-number sectors also affect $\rho_\mathrm{s}$ and other observables.

A key feature of the Luttinger-liquid phase is the power-law decay of the single-particle density matrix $\langle b_i^\dagger b_j \rangle$ defined in Eq.~\eqref{eq:spcor} in terms of the Luttinger parameter $K$. By fitting the decay as a function of the chord function $d(x,L)= L \left| \sin(\pi x/L) \right| /\pi$, we have extracted $K$ for various $t/U$, as summarized in the inset of Fig.~\ref{fig:bhphases}(a). The Luttinger parameter goes to $1$ at the Mott transition and decreases as the hopping increases. In particular, it passes through $K=1/2$ at $t/U \approx 0.24$ as will be important later on when we consider the coupling to a super-Ohmic bath.

From the density matrix, we can define the momentum distribution function
\beq
 n(k)
 = \frac{1}{L} \sum_{ij} e^{ik(i-j)} \langle b_i^\dagger b_j \rangle
\eeq 
by applying a Fourier transform with momentum $k=2\pi l/ L$, $l\in\{0, \dots, L-1\}$. Figure \ref{fig:closed_overview}(c) shows $n(k)$ for different $L$ at $t/U = 0.25$. The momentum distribution function at $k=0$ diverges with system size, but $n(k=0)/L \sim L^{-K/2}$ scales to zero in the thermodynamic limit, because particle condensation is not possible in the isolated one-dimensional system.
Furthermore, at small momenta $n(k) \sim \vert k\vert^{K/2-1}$.

From the momentum distribution function, we can define a finite-size estimator of the correlation length \cite{doi:10.1063/1.3518900},
\beq
\xi = \frac{1}{\Delta k} \sqrt{\frac{n(k=0)}{n(k=\Delta k)} - 1} \, ,
\eeq
where $\Delta k=2\pi/L$ is the resolution in momentum space.
The ratio $\xi / L$ scales to zero in the Mott phase, diverges in the long-range ordered phase and becomes scale invariant at criticality. Scale invariance also holds within the Luttinger-liquid phase, as can be checked from the $L$ dependence of $n(k)$ stated above. Figure \ref{fig:closed_overview}(d) shows $\xi/L$ across the Mott transition; $\xi/L$ scales to zero within the Mott phase and approaches a constant within the quasi-ordered phase.
In both phases, finite-size corrections lead to a convergence of $\xi/L$ from above to their values in the thermodynamic limit.
Note that the quasi-long-range-ordered phase of the classical XY model in two dimensions also has a finite $\xi/L$, which can be used to distinguish different phases \cite{PhysRevB.98.144421}.

\subsection{Ohmic Bath}

We now consider the model with a finite coupling $\alpha$ to an Ohmic bath for which the power-law exponent in Eq.~\eqref{eq:spectraldensity} is taken to be $s=1$. For this choice, the power counting arguments of Section~\ref{ss:powercounting} lead us to expect that the quasi-superfluid phase is unstable to an infinitesimal bath coupling. We now check this using QMC simulations that also allow us to determine the fate of the liquid phase. 
Moreover, we study the effects of the bath on the Mott insulator.

For all simulations, we use a cutoff frequency of $\omega_c / U =1$ for the bath. Note that the choice of $\omega_c$ does not affect the qualitative features of the phase diagram nor the critical properties, for which only the long-time decay of the retarded interaction is important. However, changing $\omega_c$ will affect the precise values of critical couplings. Because the bath is simulated exactly in terms of a retarded interaction, we do not need to apply a boson cutoff here. Moreover, we choose inverse temperatures $\beta$ that are large enough that for each $L$ our results are converged to the ground state. A detailed analysis of the temperature convergence is presented in Appendix~\ref{App:convergence_ohmic}.

\subsubsection{Dissipation-induced superfluid order}
\label{Sec:DissOrder}

\begin{figure}[t]
    \centering
    \begin{overpic}[width=0.9\linewidth]{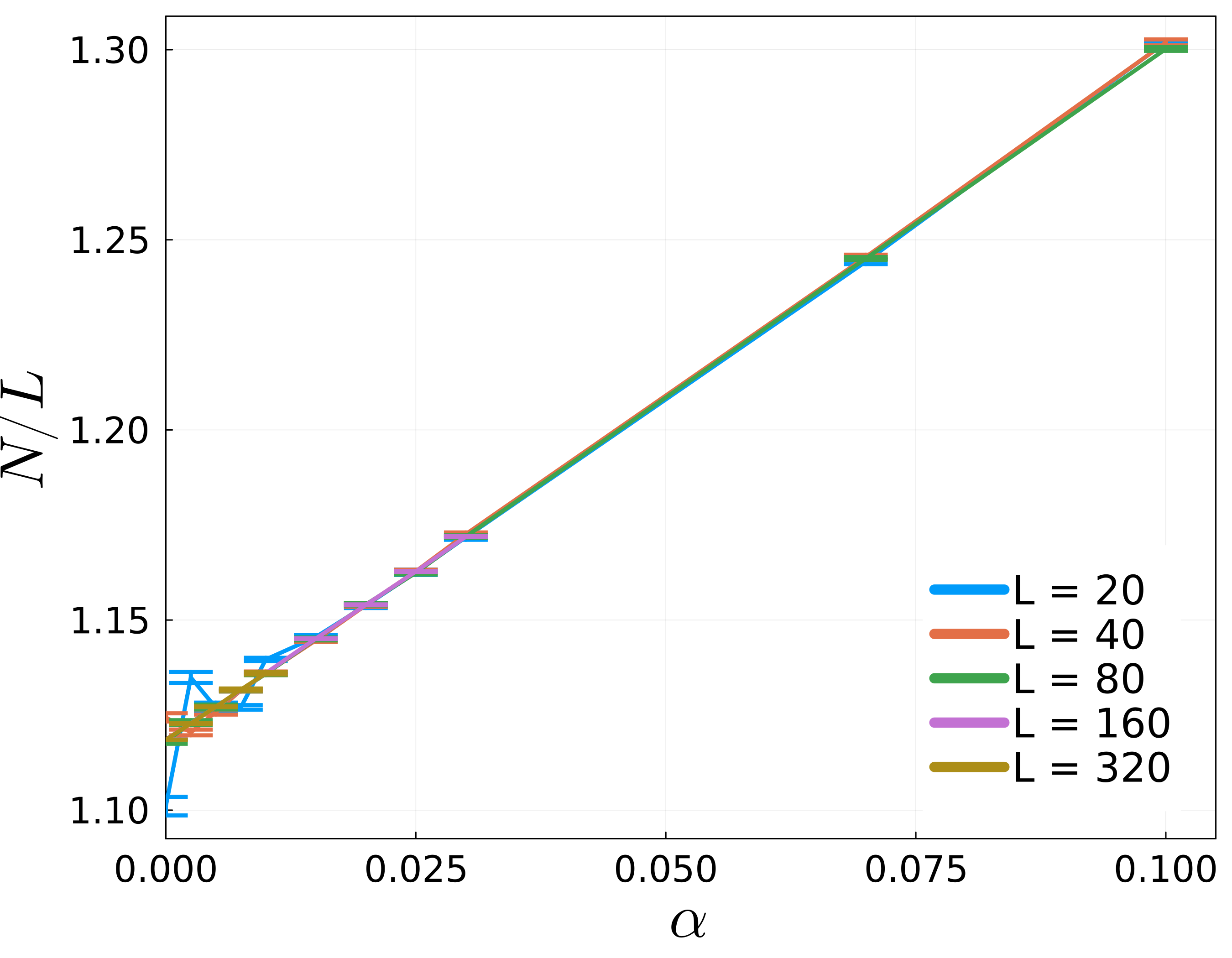}
    \put(17.5,68.5){\large(a)}
    \end{overpic}
    \begin{overpic}[width=0.9\linewidth]{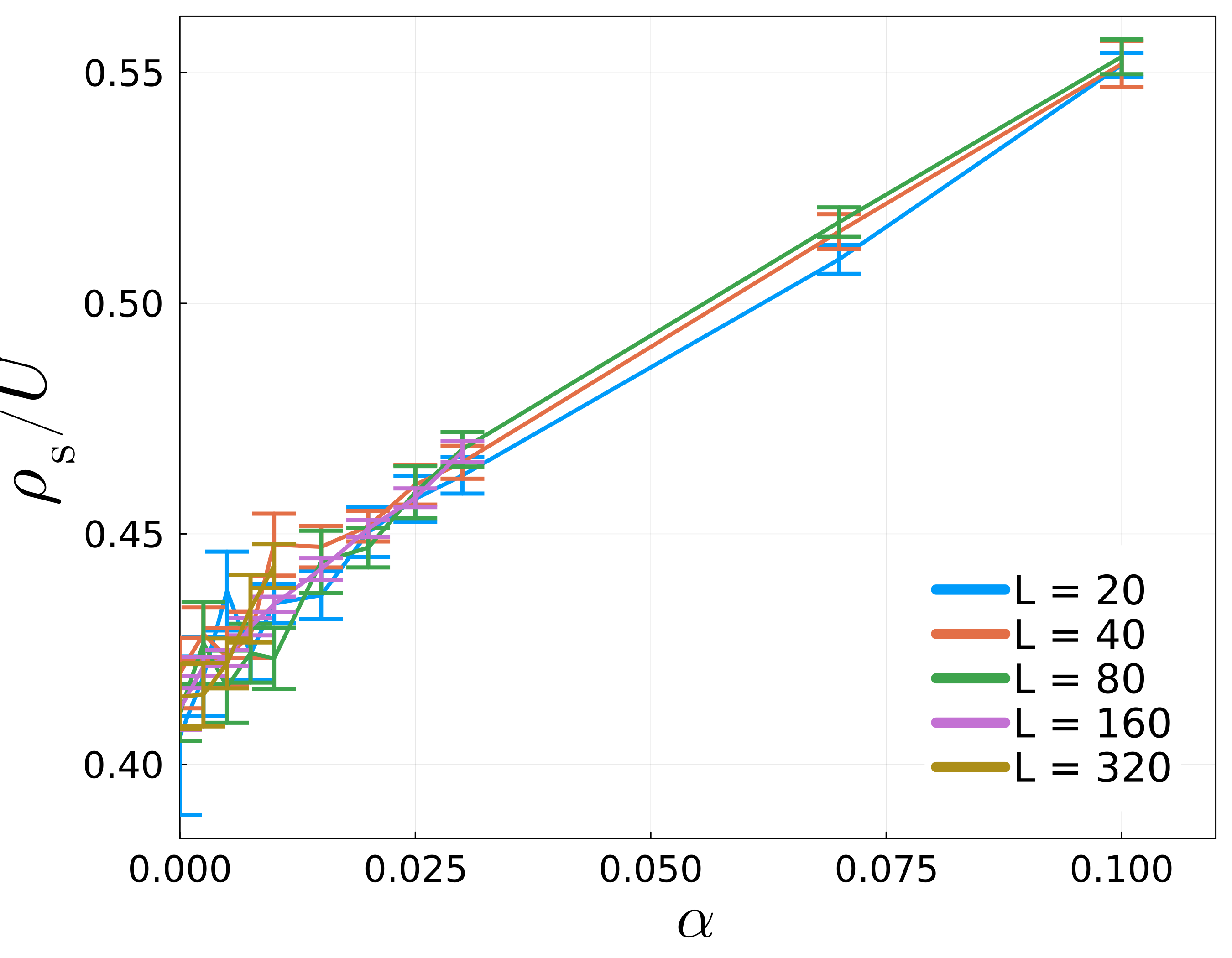}
    \put(17.5,68.5){\large(b)}
    \end{overpic}
    \caption{(a) Average particle density $N/L$ and (b) superfluid stiffness $\rho_\mathrm{s}$ as a function of the dissipation strength $\alpha$ and for various system sizes. Here, $t/U=0.25$, $\mu/U=0.3$, and  $s=1$.
    }
    \label{fig:renorm}
\end{figure}

We begin by exploring the mean particle density $N/L$ and the superfluid stiffness $\rho_\mathrm{s}$ in the chain as a function of the bath coupling for $t/U=0.25$ that puts the system in the Luttinger-liquid phase at least for zero bath coupling. Figure \ref{fig:renorm} shows these two quantities as a function of $\alpha$ for different system sizes. Within errors there is no discernible system size dependence, apart from the oscillations in $N/L$ at small $L$ and weak $\alpha$. The origin of these finite-size effects is as in the closed system, but the bath lifts the quantization of the average particle number in the ground state of our finite-size system
and quickly washes out all size effects with increasing $\alpha$ (also see Appendix~\ref{App:FSS_isolated}).
Both the particle density and superfluid stiffness can be seen to increase with increasing dissipation. At first sight, therefore, it appears that the liquid phase is stable to dissipation and that the bath merely leads to a renormalization of the parameters. We shall shortly see that the effects are in fact more dramatic than these results would suggest.

\begin{figure}[t]
    \centering
    \begin{overpic}[width=0.9\linewidth]{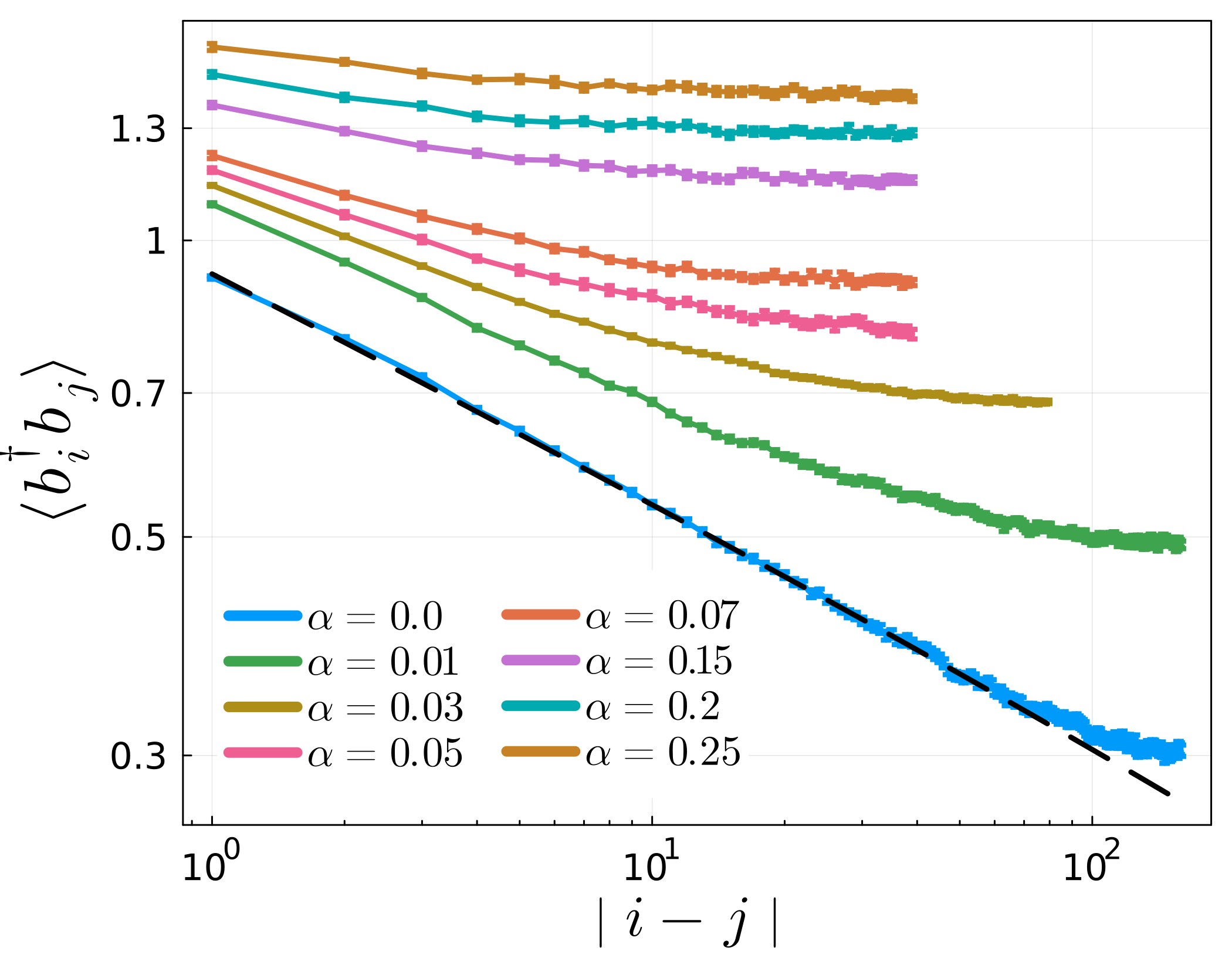}
    \put(88,68.5){\large(a)}
    \end{overpic}
    \begin{overpic}[width=0.9\linewidth]{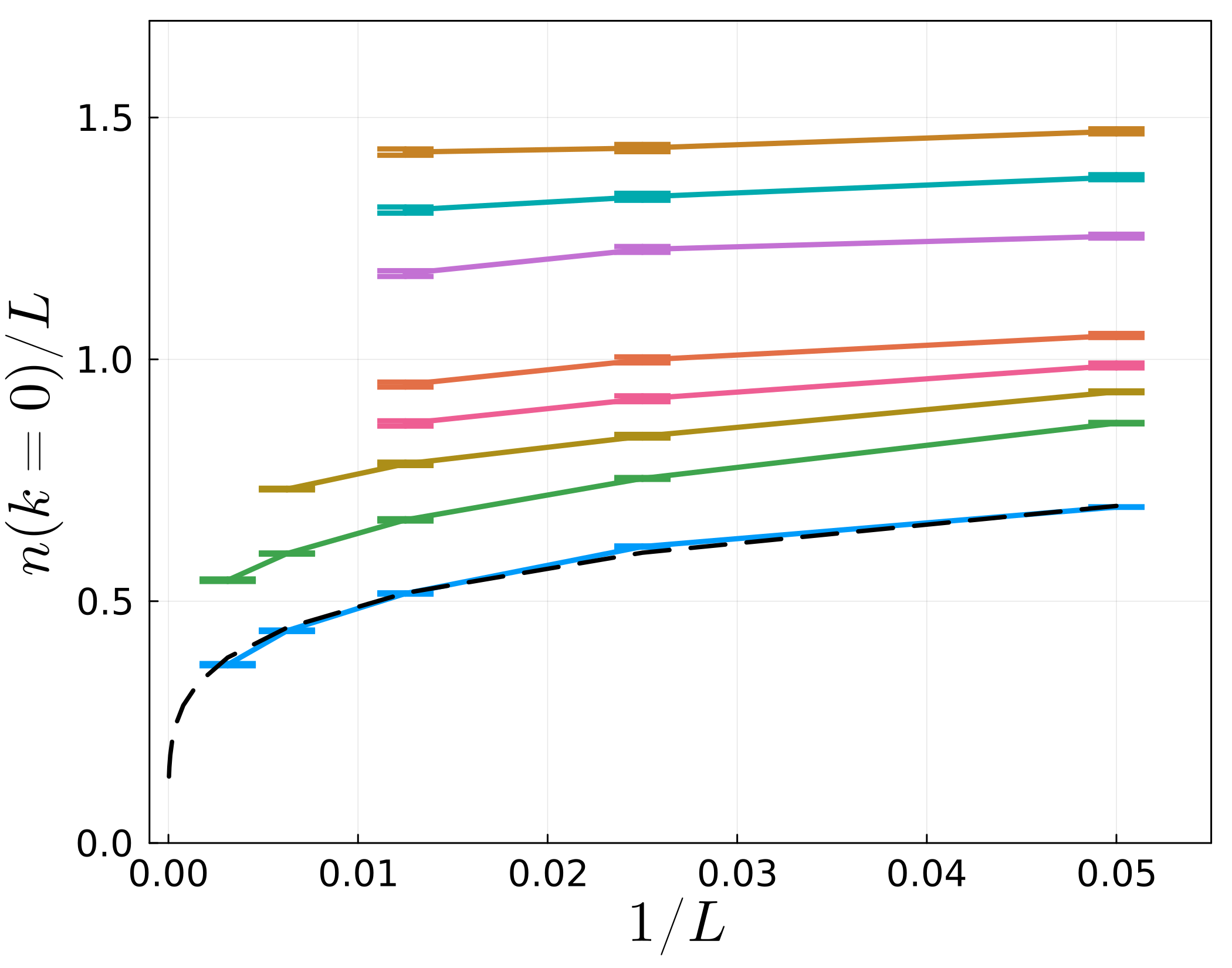}
    \put(88,15){\large(b)}
    \end{overpic}
    \caption{(a) Single-particle density matrix $\langle b^\dagger_i b_j\rangle$ as a function of distance and (b) density of the uniform mode, $n(k=0)/L$, as a function of inverse system size for different values of the dissipation strength. Note that we use a log-log scale for panel (a). The dashed black lines in (a) and (b) represent fits to the asymptotic power-law decays $\langle b_i^\dagger b_j \rangle \sim \left|i-j\right|^{-K/2}$ and $n(k=0)/L \sim L^{-K/2}$, respectively, of the non-dissipative case. 
    Here, $t/U=0.25$, $\mu/U=0.3$, and $s=1$.}
    \label{fig:dmliquid}
\end{figure}

The first glimpse that there may be more to the bath coupling than a renormalization of the parameters comes from the distance dependence of the single-particle density matrix $\langle b^\dagger_i b_j \rangle$ plotted in Fig.~\ref{fig:dmliquid}(a). For $\alpha=0$, as outlined in the previous section, the linear scaling on a log-log scale reflects the presence of the underlying Luttinger liquid. This linear scaling is, of course, cut off by the finite system size $L$. When the bath coupling is switched on, there are three main effects. One is an upward shift of the curves, another is a decrease in the initial linear slope as $\alpha$ increases and, finally, the departure from linearity drifts to smaller distances as $\alpha$ increases. The overall impression is that, at least at strong dissipation, the Luttinger liquid does not survive the introduction of a bath coupling and that there is instead a tendency to long-range order reflected in the long distance plateau for distances much less than the finite system cutoff. 

The finite-size scaling of the density of the uniform mode, $n(k=0)/L$, plotted in Fig.~\ref{fig:dmliquid}(b) shows that it falls off for larger system sizes consistent with an approach to zero in the thermodynamic limit. As $\alpha$ increases, $n(k=0)$ becomes larger for any fixed system size, though, from these results, it is not evident what happens to $n(k=0)$ as $L\rightarrow \infty$.

\begin{figure}[t]
    \centering
    \begin{overpic}[width=0.9\linewidth]{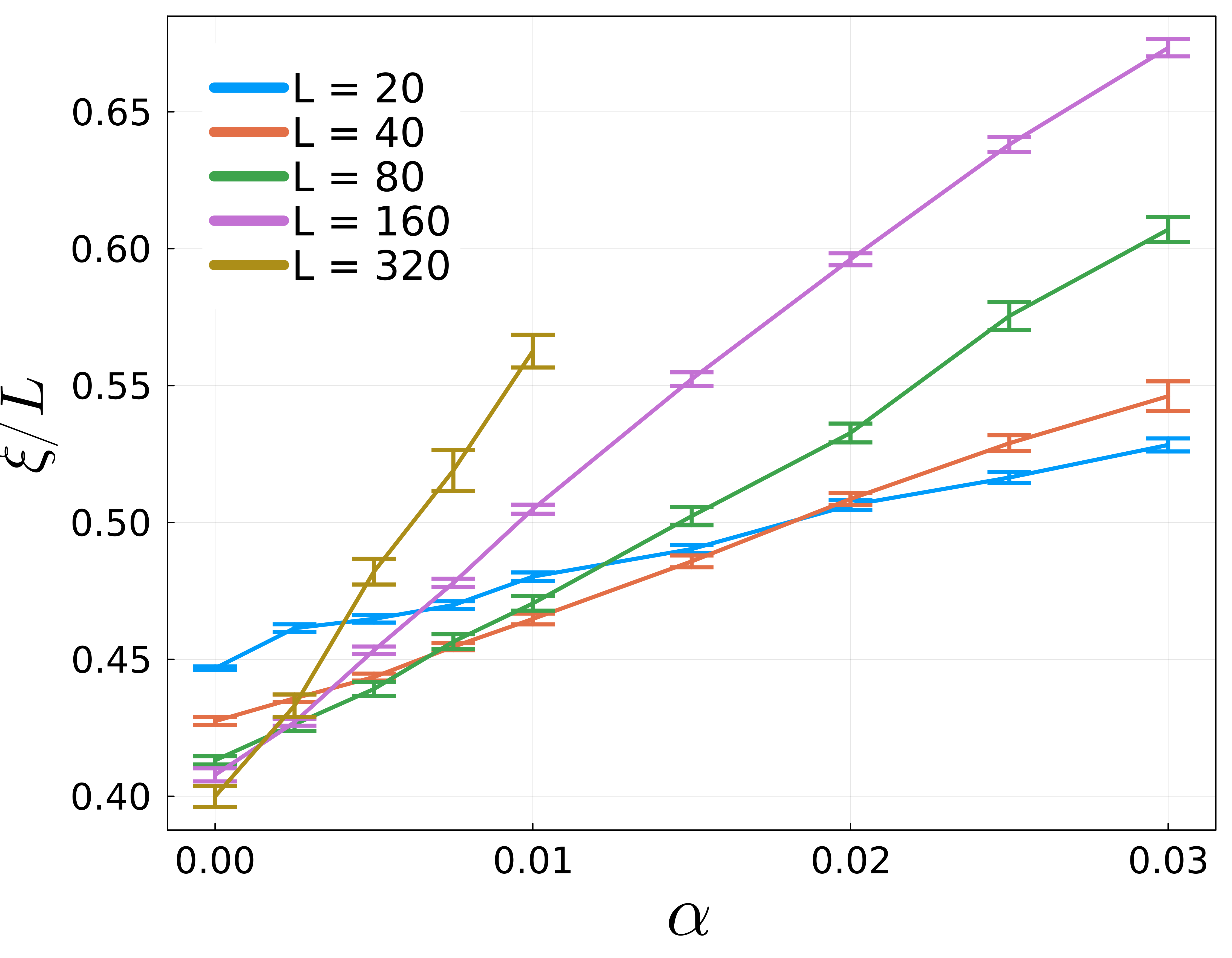}
    \put(88,15){\large(a)}
    \end{overpic}
    \\
    \hspace{-0.5cm}
    \begin{overpic}[width=0.95\linewidth]{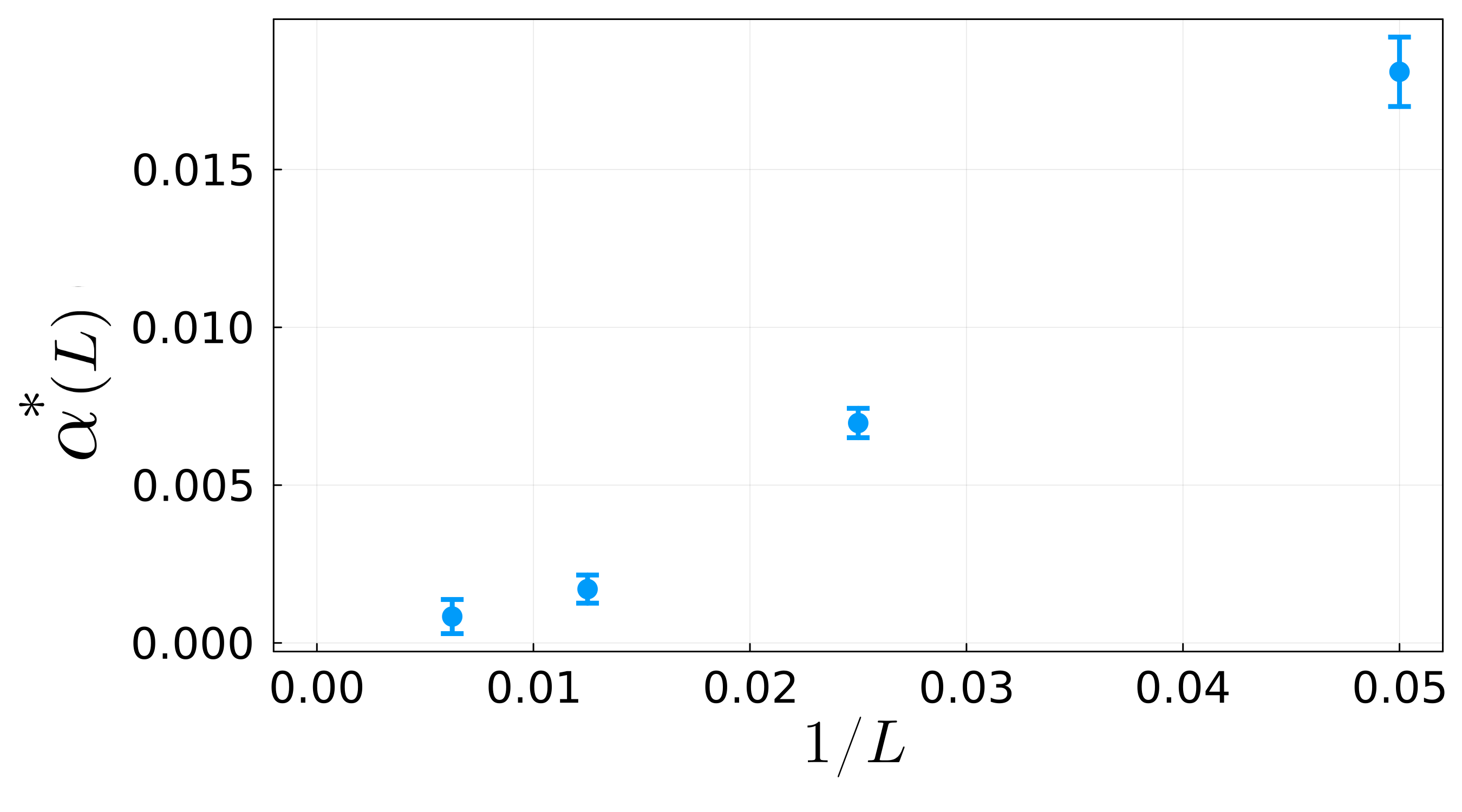}
    \put(88,15){\large(b)}
    \end{overpic}
    \caption{(a) Correlation length $\xi / L$ as a function of the dissipation strength $\alpha$ for different system sizes. (b) Finite-size extrapolation of the crossings $\alpha^\ast(L)$ between data sets $(L,2L)$ as a function of $1/L$. 
    Here $t/U=0.25$, $\mu/U=0.3$, and $s=1$.}
    \label{fig:corrcrossings}
\end{figure}

The nature of the Luttinger-liquid phase at finite dissipation becomes clearer by inspecting the normalized correlation length $\xi/L$ at zero temperature for varying $\alpha$ and for different system sizes, as shown in Fig.~\ref{fig:corrcrossings}(a). The finite-size scaling properties of $\xi/L$ are more favorable than for the quantities discussed before. Already for a rather weak bath coupling of $\alpha=0.03$, $\xi/L$ diverges with increasing $L$ and provides strong evidence for long-range order, whereas results from $\langle b^\dagger_i b_j \rangle$ and $n(k)$ are less conclusive for the same coupling.  As a function of $\alpha$, $\xi/L$ monotonically increases with a slope that becomes steadily steeper for increasing $L$. 
By contrast, in the $\alpha \to 0$ limit $\xi/L$ slowly decreases with $L$ and converges to a constant. As a result, $\xi / L$ lines for data pairs $(L,2L)$ exhibit a crossing at 
 a finite $\alpha^\ast(L)$.
In case of a finite-$\alpha$ continuous transition and in the absence of finite-size corrections, there would be a crossing of all the lines at a common value of $\alpha$. Instead, we find that the crossing of the lines drifts with $L$. To get reliable estimates for $\alpha^*(L)$, we have performed polynomial fits close to the crossings and obtained the error from a bootstrap analysis. 
Our results are collected in Fig.~\ref{fig:corrcrossings}(b) which shows that $\alpha^*(L)$ tends to zero as $1/L \rightarrow 0$. This finite-size scaling result indicates that the Luttinger liquid is destabilized for any finite bath coupling in agreement with the power-counting argument. The resulting phase appears to be long-range ordered for any $\alpha > 0$.

The central result in Fig.~\ref{fig:corrcrossings} provides information about the thermodynamic limit of the model at zero temperature. In order to make statements about the ground state, it was necessary to achieve convergence of $\xi/L$ by tuning the temperature, as analyzed in detail in Appendix~\ref{App:convergence_ohmic}. One could have also performed a finite-size scaling at fixed $\beta / L^z$, provided that we know the dynamical critical exponent $z$. In field theory, the free propagator is proportional to $1/(k^2 + c_1 \, \omega^2 + c_2 \, \omega^s)$ suggesting $z=2$ for $s=1$.
At large dissipation strengths, the temperature convergence discussed in Appendix~\ref{App:convergence_ohmic} is consistent with this scenario, but not precise enough to allow us to draw a definitive conclusion. In this context it is worth remarking that $z=2$ was found for the transition from the disordered to the ordered phase of the dissipative O(2) quantum rotor model in one dimension \cite{2005JPSJ...74S..67W,PhysRevB.85.214302} and from spin-wave theory \cite{Weber2022}.

The summary of this part is that the Luttinger liquid phase of the Bose-Hubbard model in one dimension is unstable to coupling to an Ohmic bath. Fluctuations, that would ordinarily destroy long-range order in one dimension, are suppressed by the bath coupling through the appearance of a long-range effective potential in the imaginary time direction.

\subsubsection{Fate of the Mott insulator}
\label{Sec:Mott}

We now turn our attention to the Mott phase coupled to the bath. Since this is a gapped phase of matter, it is stable to sufficiently weak Hamiltonian couplings between the bosonic degrees of freedom on the chain. But here we are coupling to an external system that is gapless so the nature of the resulting state of matter requires some investigation. 

\begin{figure}[t]
    \centering
    \begin{overpic}[width=0.9\linewidth]{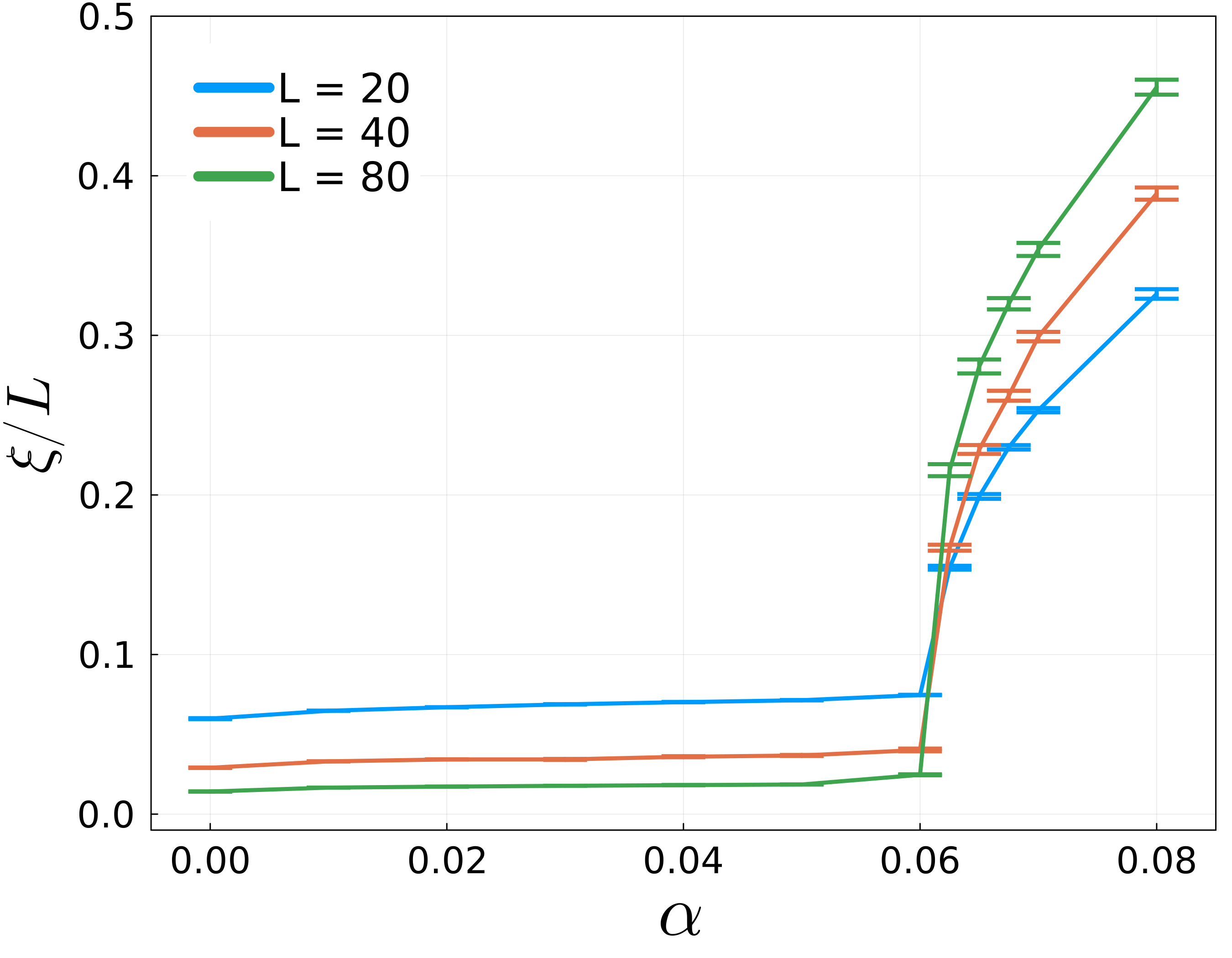}
    \put(88,15){\large(a)}
    \end{overpic}
    \\ \hspace{-0.3cm}
    \begin{overpic}[width=0.92\linewidth]{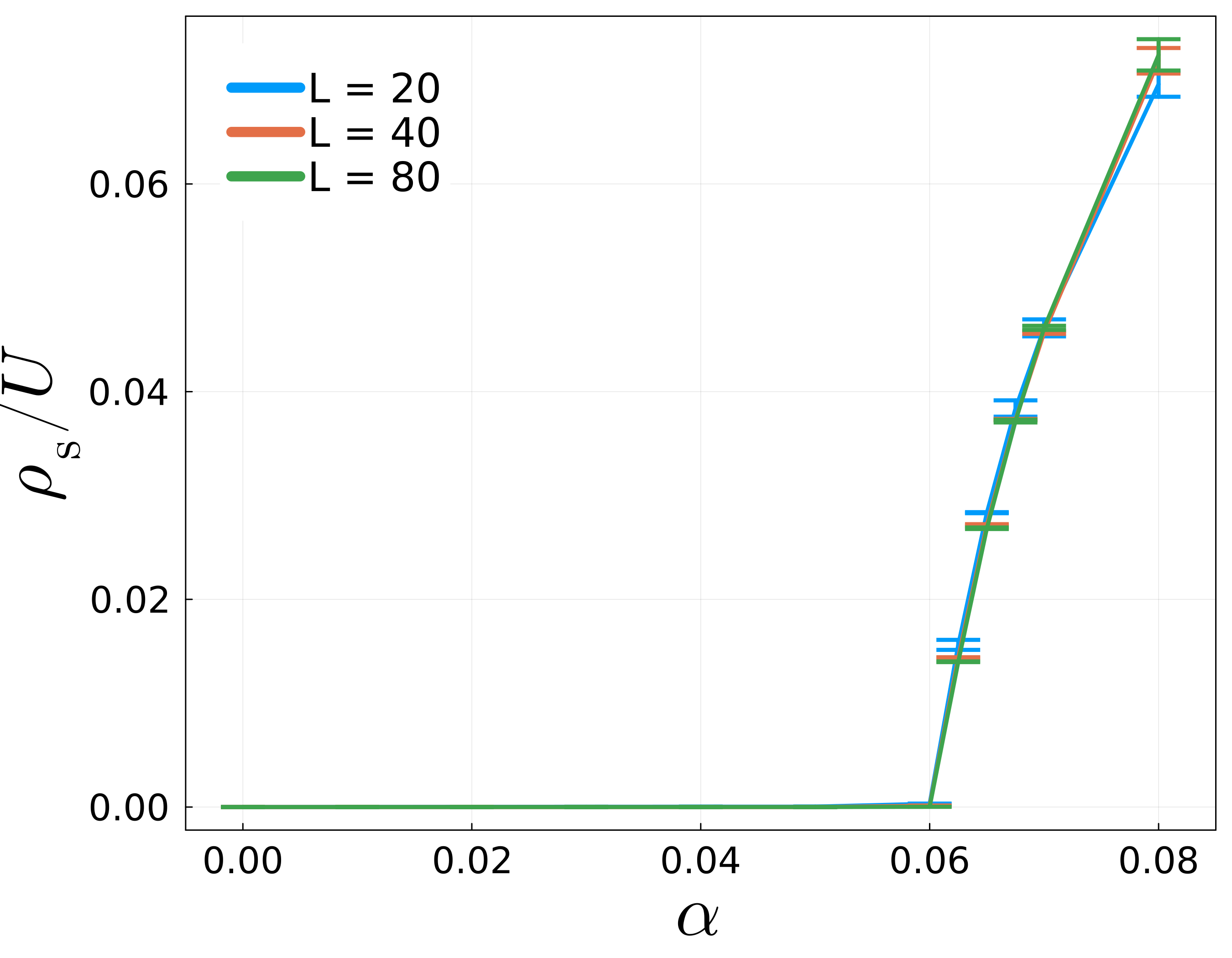}
    \put(88,15){\large(b)}
    \end{overpic}
    \caption{(a) Correlation length $\xi/L$ and (b) the superfluid stiffness $\rho_\mathrm{s}$ as a function of $\alpha$ for different system sizes. Here, $\mu/U = 0.3$, $t/U = 0.1$, and $s=1$.}
    \label{fig:mott2liquid}
\end{figure}

Figure \ref{fig:mott2liquid}(a) shows how $\xi/L$ varies with $\alpha$ for different system sizes. This clearly shows two regimes separated by a transition at $\alpha_c\approx0.061$. For $\alpha < \alpha_c$, $\xi/L$ tends towards zero as the thermodynamic limit is approached while the correlation length grows faster than the system size for $\alpha>\alpha_c$. The data shows a common crossing that identifies the transition. The trend in $\xi/L$ with $\alpha$ is reminiscent of the behaviour of the system starting from the liquid phase. It is reasonable to expect therefore that the small-$t/U$ large-$\alpha$ phase is a long-range ordered phase that is adiabatically connected to the large-$t/U$ phase at finite $\alpha$. A distinction between the two small-$t/U$ phases is seen also in the superfluid stiffness [Fig.~\ref{fig:mott2liquid}(b)] which is zero for all $\alpha < \alpha_c$ and non-vanishing for larger bath coupling. Again, the temperature convergence of $\xi/L$ across the transition is discussed in Appendix~\ref{App:convergence_ohmic}.

\begin{figure}[t]
    \centering
    \begin{overpic}[width=0.9\linewidth]{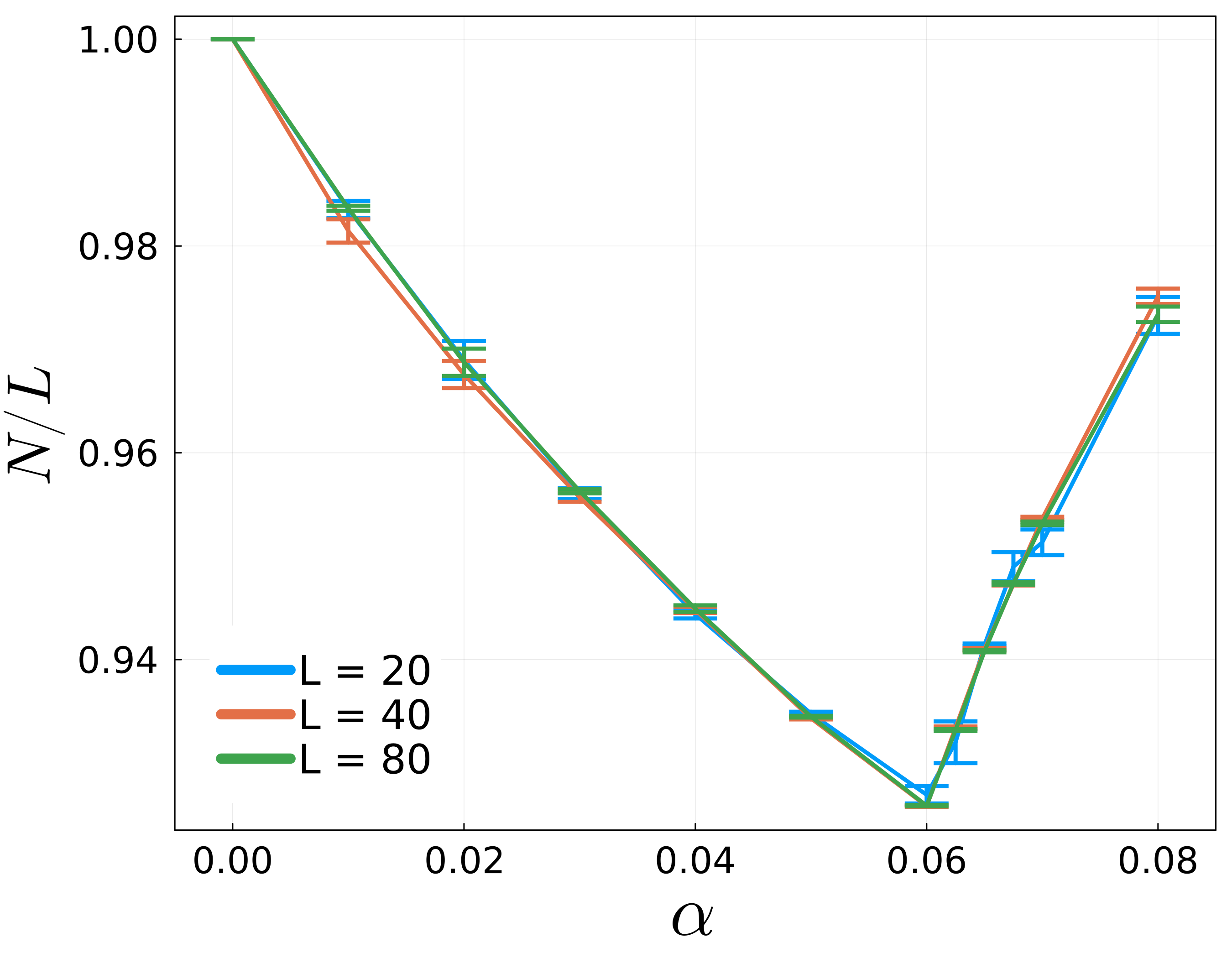}
    \put(88,15){\large(a)}
    \end{overpic}
    \begin{overpic}[width=0.9\linewidth]{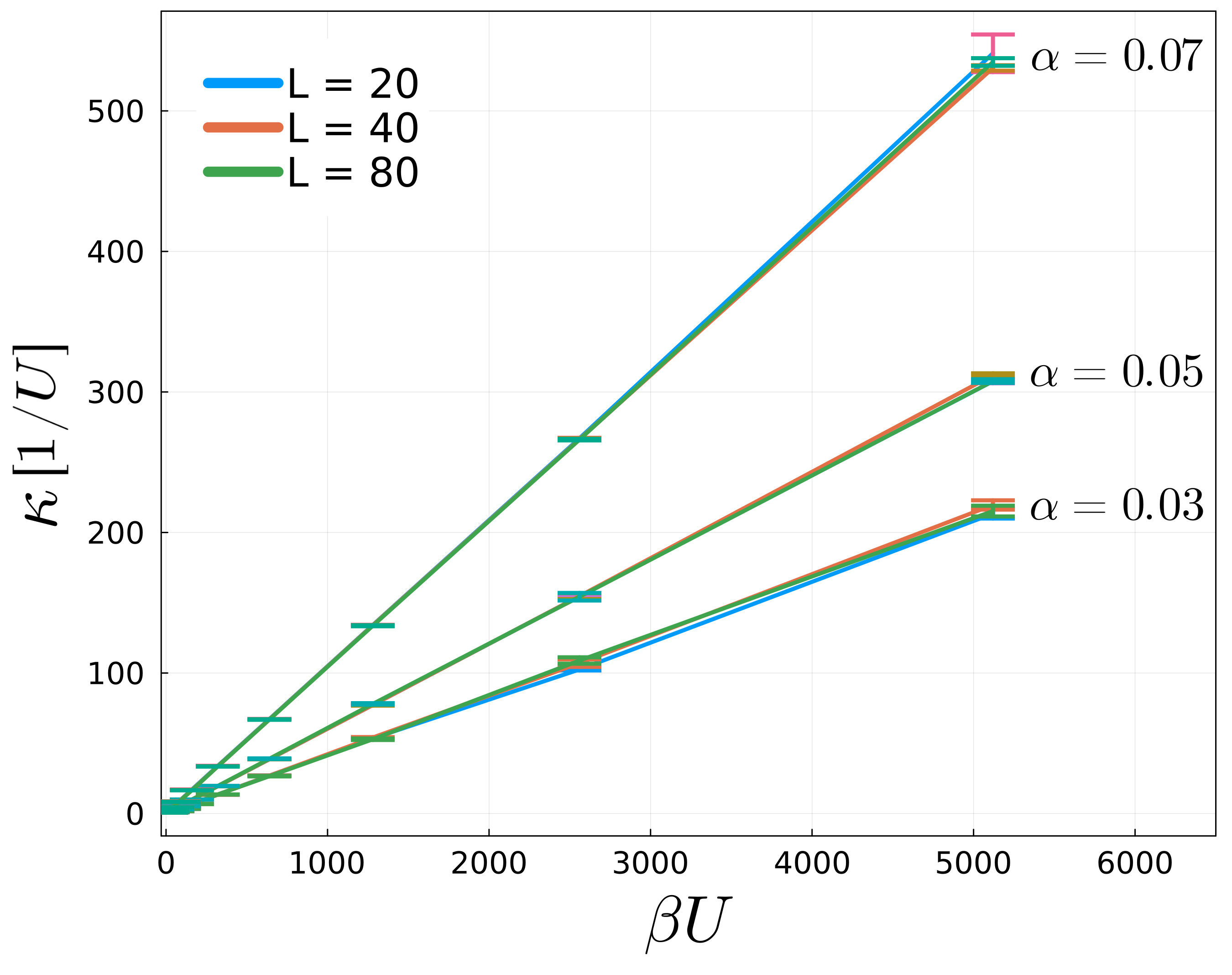}
    \put(88,15){\large(b)}
    \end{overpic}
    \caption{(a) Average local boson occupation, $N/L$, as a function of the dissipation strength $\alpha$ for different system sizes. (b) Compressibility $\kappa$ as a function of inverse temperature $\beta$ for different $\alpha$.
    Here, $\mu/U=0.3$, $t/U=0.1$, and $s=1$.}
    \label{fig:compressibility}
\end{figure}

The vanishing superfluid stiffness and correlation length are both consistent with a Mott phase. However, the dissipative coupling causes the system to depart significantly from what we would ordinarily expect of such a phase. This becomes clear from the compressibility computed from the particle-number fluctuations of the system, \ie,
\beq
\kappa = \frac{\beta}{L}\left( \langle N^2 \rangle - \langle N \rangle^2 \right).
\label{eq:compressibility}
\eeq
The average local particle number $N/L$ and the compressibility $\kappa$ are shown in Fig.~\ref{fig:compressibility}.
The Mott phase (at zero $\alpha$) is incompressible and the particle number per site is strictly one. Our QMC results are consistent with this. However, when the bath coupling is switched on, $N/L$ decreases and the particle number acquires a variance that quickly saturates with system size, $L$, and inverse temperature $\beta$. This implies [Eq.~\eqref{eq:compressibility}] that the compressibility diverges at zero temperature in the dissipative Mott phase, consistent with previous work on particle dissipation \cite{PhysRevB.97.035148}. The compressibility also diverges in the dissipation-induced long-range ordered phase, as illustrated for $\alpha=0.07$ in Fig.~\ref{fig:compressibility}(b). 
In general, our numerical results suggest that the compressibility of the system diverges for any finite $\alpha$, independent of the bath exponent $s$ and the phases we are in. This is a direct consequence of particle exchange between our system and the gapless bath which has always low-energy states available. As mentioned above we call the resulting state at finite bath coupling Mott$^{\ast}$.

Interestingly, the effect of dissipation on the total particle number is opposite in the two phases of the Bose-Hubbard model: starting from the Luttinger-liquid phase, particle dissipation adds bosons to the system, whereas it removes particles from the Mott phase. The fact that the boson number is not pinned to integer fillings anymore in the insulating phase is a consequence of the nonzero commutator $[H,\sum_i b^\dagger_i b_i] \neq 0$ in the presence of the bath.

We have repeated our analysis for other values of the hopping $t/U$. Results for the phase boundaries are collected in the phase diagram depicted in Fig.~\ref{fig:bhphases}(b).

\subsection{Super-Ohmic bath}

In accordance with the power-counting argument presented in Section~\ref{ss:powercounting}, we have found that the Ohmic bath with $s=1$ is a relevant perturbation throughout the entire Luttinger-liquid phase. We have also seen that the resulting phase is a long-range-ordered superfluid state. In the following, we tune the bath exponent into the super-Ohmic regime $s>1$ with the expectation that the Luttinger liquid is stable within parts of the phase diagram.

\begin{figure}[t]
    \centering
    \begin{overpic}[width=0.9\linewidth]{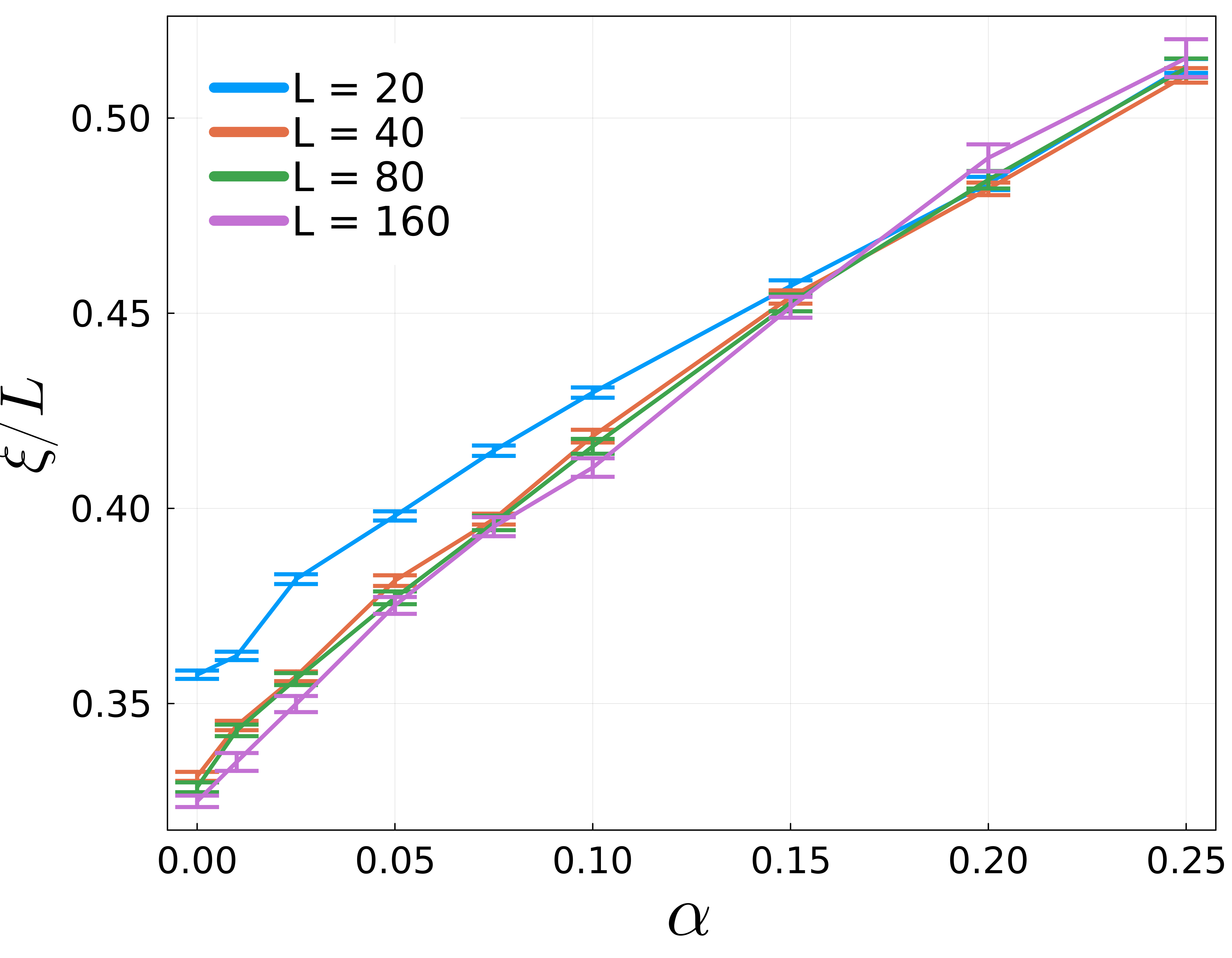}
    \put(88,15){\large(a)}
    \end{overpic}
    \begin{overpic}[width=0.9\linewidth]{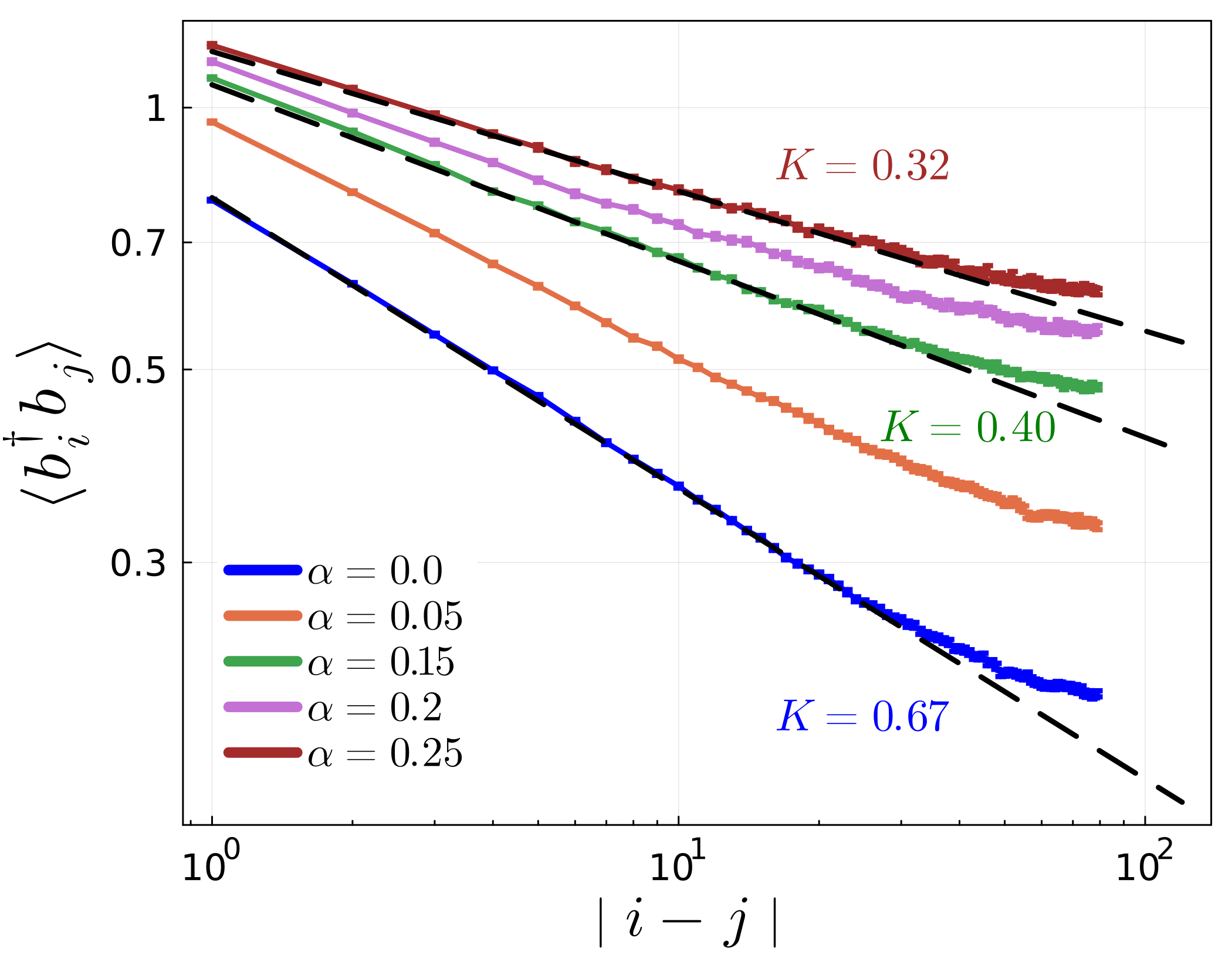}
    \put(88,67.5){\large(b)}
    \end{overpic}
    \caption{Plot illustrating the stability of the Luttinger liquid at $s=2.5$. (a) Correlation length $\xi/L$ as a function of $\alpha$ for different system sizes. (b) Single-particle density matrix $\langle b^\dagger_i b_j \rangle$ as a function of distance. We performed power-law fits to quantify the change of the Luttinger parameter $K$ as $\alpha$ is varied. Here, $\mu/U=0.5$ and $t/U= 0.16$.}
    \label{fig:s25}
\end{figure}

\subsubsection{Renormalization of the Luttinger liquid at $s=2.5$}

We first consider the coupling to a bath with exponent $s=2.5$, which is an irrelevant perturbation throughout the entire Luttinger-liquid phase. Figure \ref{fig:s25}(a) shows the correlation length $\xi/L$ as a function of $\alpha$ at $\mu/U = 0.5$ and $t/U=0.16$. Apart from the smallest system size, all data sets collapse on top of each other, indicating that the Luttinger liquid remains stable. This is also confirmed by the density matrix $\langle b^\dagger_i b_j\rangle$ depicted in Fig.~\ref{fig:s25}(b). 
In contrast to Fig.~\ref{fig:dmliquid}(a), the power-law decay of $\langle b^\dagger_i b_j\rangle$ remains clearly visible up to the largest couplings; the final upturn is caused by boundary effects and occurs on the same length scale for all $\alpha$.
We observe that the Luttinger exponent $K$ decreases with increasing $\alpha$, as quantified by the power-law fits of $K$ stated in Fig.~\ref{fig:s25}(b). It appears that the effect of the bath is to renormalize the properties of the Luttinger liquid. It is not clear whether the bath will induce a transition at larger couplings \cite{PhysRevB.85.214302}.

\subsubsection{Transition from Luttinger liquid to superfluid at $s=1.75$}

\begin{figure}[t]
    \centering
    \includegraphics[width=\linewidth,trim={0 30 0 100},clip]{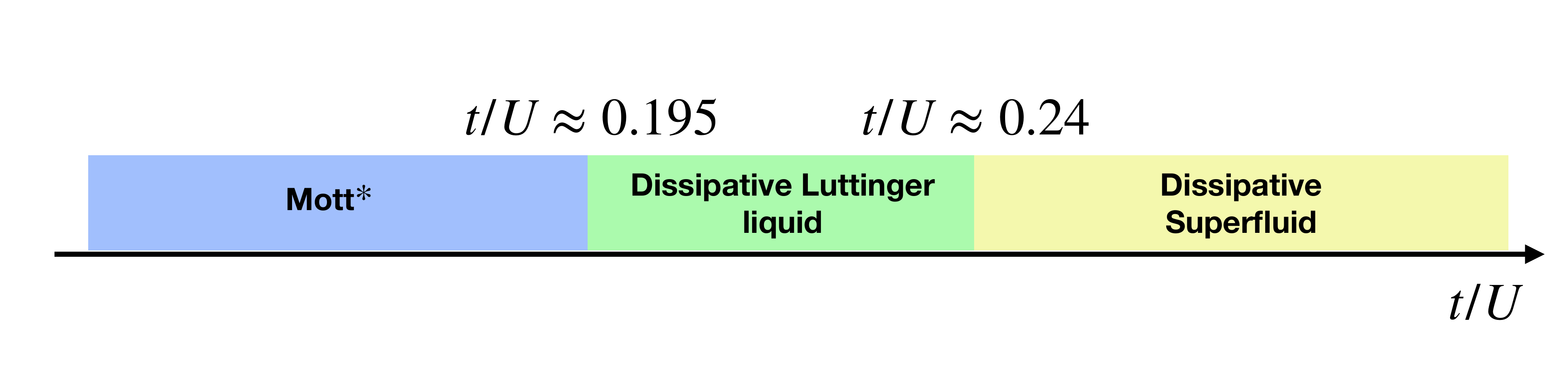}
    \caption{Expected phase diagram for $s=1.75$ and infinitesimal bath coupling $\alpha$ at $\mu / U = 0.3$ and as a function of $t/U$, as obtained from the power-counting argument presented in Section~\ref{ss:powercounting}.
    }
    \label{fig:schematic}
\end{figure}

We now study the effect of a bath with exponent $s=1.75$ at $\mu/U =0.3$. Our power-counting analysis from Section~\ref{ss:powercounting} suggests that there should be three distinct regimes at weak dissipation, as illustrated in Fig.~\ref{fig:schematic}. For $t/U \lesssim 0.195$, the Mott phase leads to a dissipative insulating phase with infinite compressibility, as discussed in Sec.~\ref{Sec:Mott}. For $t/U \gtrsim 0.24$, the bath is a relevant perturbation and we expect to find a long-range ordered phase, as in the Ohmic case of Sec.~\ref{Sec:DissOrder}. In between, power counting suggests that the Luttinger liquid remains stable for small couplings.

\begin{figure}
    \centering
    \begin{overpic}[width=0.9\linewidth]{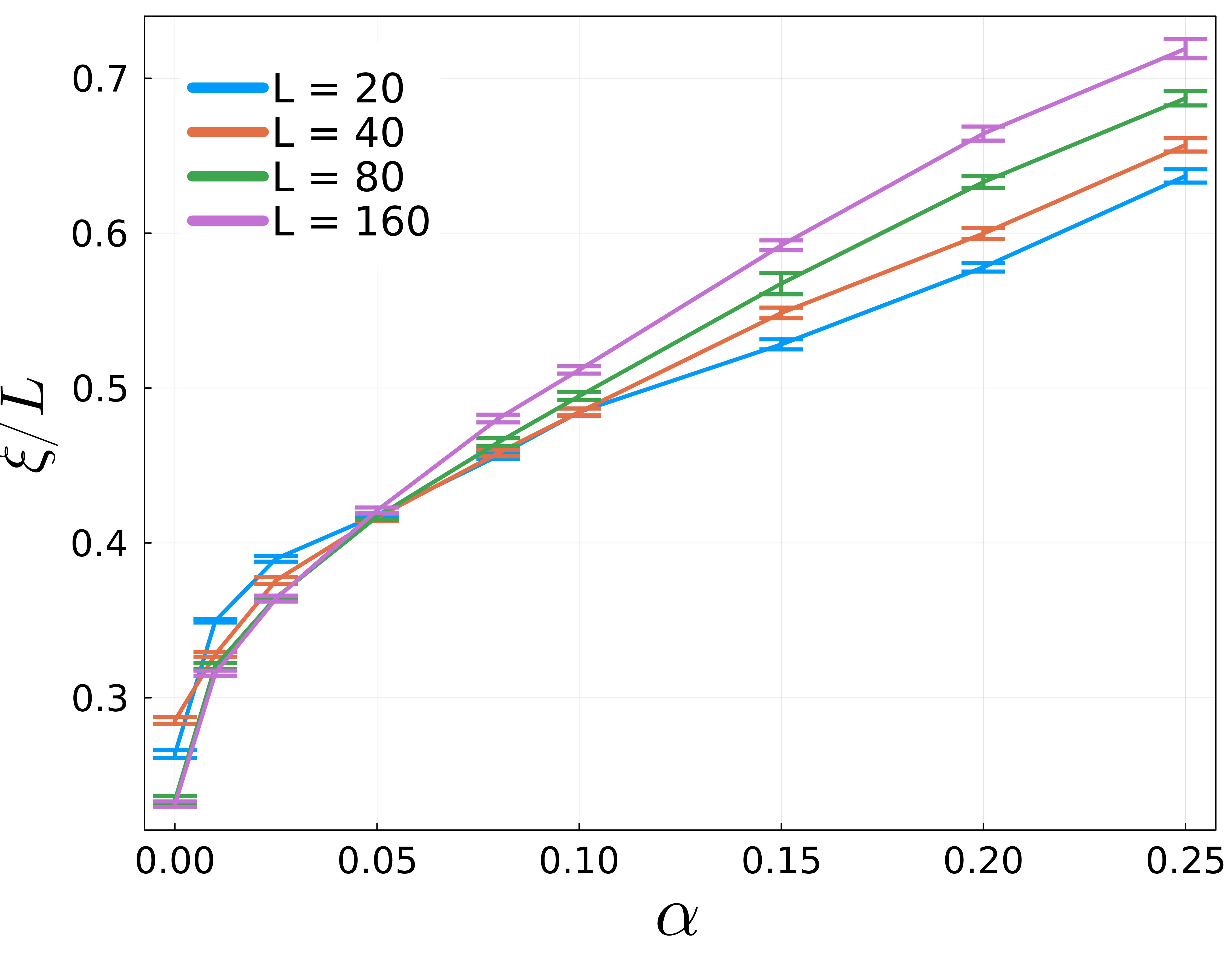}
    \put(88,15){\large(a)}
    \end{overpic}
    \\ \hspace{-0.3cm}
    \begin{overpic}[width=0.94\linewidth]{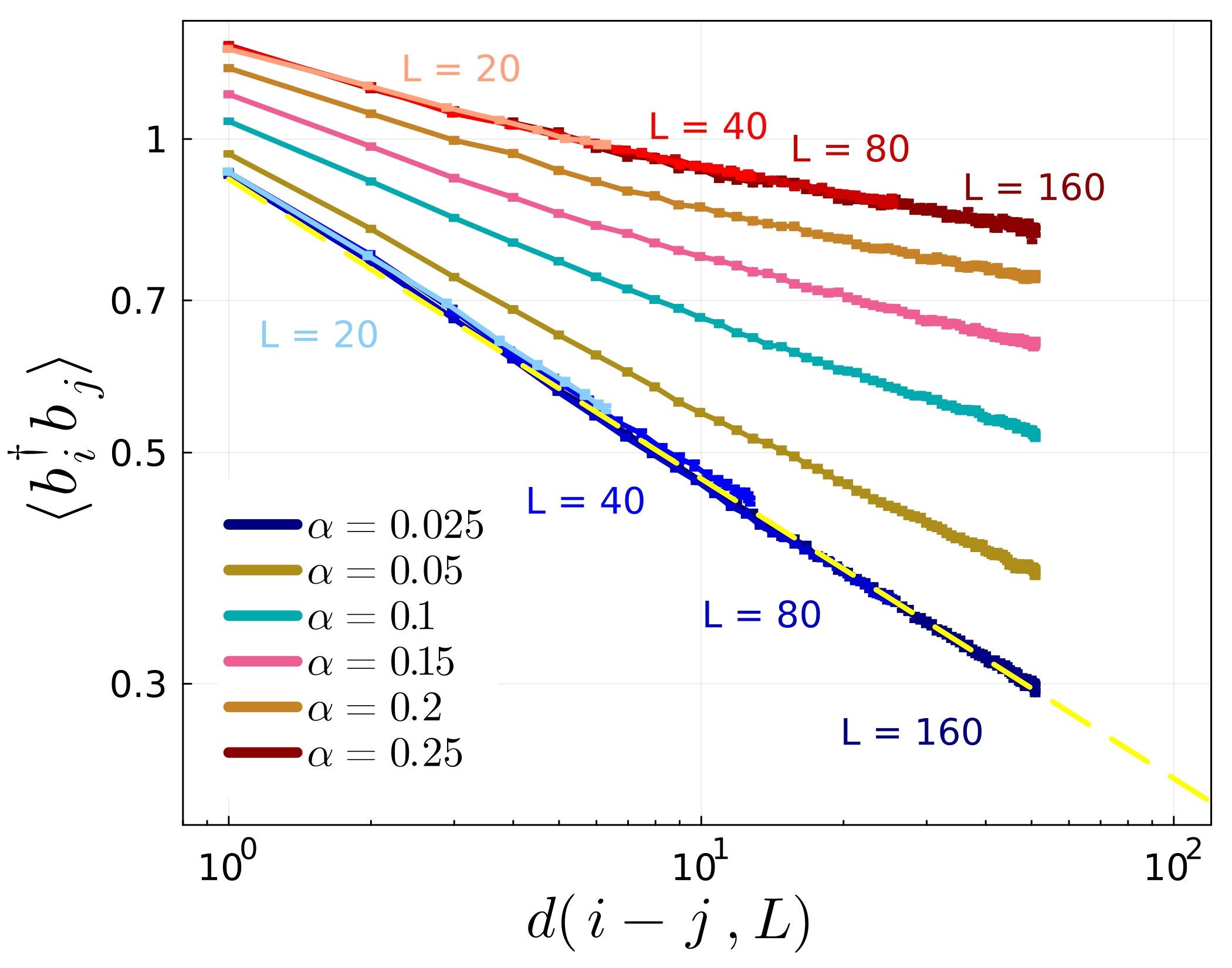}
    \put(88,69.5){\large(b)}
    \end{overpic}
    \caption{Putative transition between the Luttinger liquid and the long-range ordered phase at $s=1.75$, $\mu/U=0.3$, and $t/U=0.2$, as illustrated with (a) the correlation length $\xi/L$ as a function of $\alpha$ and (b) the density matrix $\langle b^\dagger_i b_j \rangle$ as a function of distance. Here we use the chord function $d(x,L)= L \left| \sin(\pi x/L) \right| /\pi$ to get rid of the boundary effects within the Luttinger-liquid phase, to better compare with the power-law fit for $\alpha=0.025$. For the weakest and strongest couplings, we also show how different system sizes collapse onto the same curve. }
    \label{fig:s175}
\end{figure}

Figure \ref{fig:s175}(a) shows the correlation length $\xi/L$ as a function of the dissipation strength $\alpha$ in the intermediate hopping regime at $t/U = 0.2$. We choose a value of the hopping that is close to the Mott transition, so that the Luttinger parameter of the isolated chain is close to one [see Fig.~\ref{fig:bhphases}(a)], which maximizes the absolute value of the scaling dimension in Eq.~\eqref{eq:scalingdim}. At strong $\alpha$, we find a clear separation between data sets of $\xi/L$ at different lattice sizes which slowly increases with $L$. This behavior is reminiscent of the emergence of long-range order found for the Ohmic case. In contrast to our analysis in Fig.~\ref{fig:corrcrossings}, the larger bath exponent $s$ leads to a much slower divergence of the different curves. At weak coupling, $\xi/L$ shows no visible deviations between $L=80$ and $L=160$ for $\alpha < 0.05$. In accordance with our results at $s=2.5$, this can be interpreted as a signature of a stable Luttinger-liquid phase. It appears that the different curves start to depart from each other at $\alpha \gtrsim 0.05$. However, it is worth noting that even with $K \to 1$, the scaling dimension $\Delta=-0.25$ is rather small and we have to expect significant finite-size corrections. If we tune $t/U$ closer to the marginal point at $t/U \approx 0.24$ and even beyond, the even smaller scaling dimension leads to an extremely slow renormalization-group flow which makes the qualitative features at weak coupling look the same over an extended parameter range, even for $t/U=0.3$ where long-range order is supposed to appear for any $\alpha>0$ (not shown). Because the Luttinger parameter only decreases slowly with increasing $K$, we would need to simulate at extremely large $t/U$ to recover a clear signature of order at $\alpha>0$. However, for larger couplings we always see ordering tendencies, as is the case for the Ohmic bath presented in Fig.~\ref{fig:corrcrossings}.

We also analyze the real-space decay of $\langle b^\dagger_i b_j \rangle$ in Fig.~\ref{fig:s175}(b). This time, we plot the density matrix against $d(i-j,L)$ to remove boundary effects within the putative Luttinger-liquid phase. At $\alpha=0.025$, our results are consistent with a power-law decay. We also confirm that the density matrix converges to this power law, as we increase the system size from $L=20$ to $L=160$. At dissipation strengths beyond $\alpha \approx 0.05$, $\langle b^\dagger_i b_j \rangle$ gets enhanced and starts to bend upwards and deviate from the power-law towards a behavior that is consistent with long-range order.

All in all, we find numerical evidence for long-range order even at $s=1.75$ and a weak-coupling behavior that is consistent with a stable Luttinger-liquid phase. However, the weak-coupling analysis at $s=1.75$ is complicated by slow scaling and therefore has to be taken with a grain of salt.

\section{Summary and Conclusions\label{Sec:Conclusions}}

In this work we have provided a systematic, numerically exact QMC study of the one-dimensional Bose-Hubbard model coupled off-diagonally to an oscillator bath with total boson number conservation thus introducing a mechanism for dissipation and decoherence resulting in the phase diagram shown in Fig.~\ref{fig:bhphases}(b). We have investigated the fate of the Mott and quasi-superfluid phases in the zero-temperature limit as a function of the coupling to the bath. The principal finding is that, for an Ohmic bath, the quasi-superfluid or Luttinger liquid phase is unstable to the presence of infinitesimal bath coupling leading to a long-range ordered superfluid. This is possible in this one-dimensional system, in apparent violation of the Mermin-Wagner theorem, because the bath can be thought of as introducing long-range interactions in imaginary time that severely suppress fluctuation effects. In other words, the bath coupling qualitatively changes the nature of the collective physics. 
Something similar can be said for the Mott phase though here the effect is much less dramatic. Whereas the Mott phase survives in some form (zero superfluid stiffness and small correlation length), the coupling to a gapless bath generates an infinite compressibility at zero temperature. 

We have also studied super-Ohmic baths. For larger bath exponent $s=2.5$ where power counting predicts that the bath coupling is irrelevant, we find results consistent with the presence of a renormalized Luttinger liquid over the range of bath couplings explored ($\alpha \leq 0.25$) with a Luttinger parameter that steadily decreases as the bath coupling increases.
For intermediate $s=1.75$, power counting predicts three regimes: the Mott$^*$ phase, a dissipative Luttinger liquid and, for sufficiently large $t/U$, the long-range ordered superfluid. Our QMC results for this case are much less compelling than our previous results. In particular, for $t/U=0.2$ in the regime where the dissipative Luttinger liquid is expected for small bath couplings, they fairly clearly indicate long-range order for larger $\alpha$ while for smaller $\alpha$ they are merely consistent with finite $\alpha$ stability of the liquid phase.  

The existence of a long-range ordered superfluid phase in the one-dimensional dissipative Bose-Hubbard model
is in close analogy to recent work on quantum spin chains, for which an Ohmic bath spontaneously breaks the continuous spin-rotational symmetry of system plus bath and thereby induces long-range antiferromagnetic order \cite{2005JPSJ...74S..67W,Weber2022}.
A previous study of particle dissipation for a model of interacting hard core bosons did not find long-range order \cite{PhysRevB.97.035148}, which is likely a result of insufficient finite-size scaling at fixed $\beta/ L$.
Our finding of a Mott phase with diverging compressibility is in accordance with results on the dissipative charge-density-wave phase in Ref.~\cite{PhysRevB.97.035148}.
The long-range ordered superfluid phase in the Bose-Hubbard model
bears strong resemblance to the
incoherent transverse quantum fluid put forward in Ref.~\cite{Kuklov2023}
which describes long-range order in a quasi-one-dimensional open quantum system. In this scenario, an infinte compressibility leads to  a quadratical energy-momentum relation
consistent with the expected $z=2$ exponent for an Ohmic bath deep within the dissipative superfluid regime.
We leave for the future a more detailed investigation of any further connections to this phase.

The QMC technique, with wormhole updates, can be of use to study a number of dissipative quantum many-particle systems \cite{PhysRevB.105.165129}. Our extension to the worm algorithm can be of advantage in the presence of large onsite interactions, as is the case for the Bose-Hubbard model in the Mott phase. Our formulation can also be applied to Dicke-type light-matter interactions in cavities \cite{PhysRevB.105.165129}.
In future, it will be interesting to characterize the dissipative phases of the Bose-Hubbard model in more detail and compare different dissipation mechanisms such as the coupling to the local density, that was studied using mean-field theory \cite{PhysRevA.79.053611} or QMC for hard core bosons \cite{Cai2014}. 
Moreover, the dynamical properties of the dissipation-induced superfluid phase would be good to characterize. 
The interplay between dissipation and additional interaction terms motivates further studies of these systems. It is known that in the presence of disorder the Mott lobes become bounded by a so-called Bose glass phase \cite{POLLET2013712}. The fate of this phase in the presence of dissipation might be feasible using the methods discussed here. 
\\

\textit{Note added:} Upon completion of this work, we became aware of Ref.~\cite{kuklov2023universal} where signatures of the dissipation-induced long-range ordered phase have been detected for the hard core boson model of Ref.~\cite{PhysRevB.97.035148} at temperatures $\beta \propto L$ using analytical predictions based on the theory for incoherent transverse quantum fluids. This is consistent with our direct numerical evidence of the superfluid phase for Ohmic dissipation.

\begin{acknowledgments}
We acknowledge use of the computing resources of the Max Planck Institute for the Physics of Complex Systems.
PR and AR acknowledge partial support from Fundação para a Ciência e Tecnologia (FCT-Portugal) through Grant No. UID/CTM/04540/2020. PR acknowledges partial support from Horizon 2020 - QuantERA II Programme under Grant Agreement No.\ 101017733 \footnote{\url{https://doi.org/10.54499/QuantERA/0003/2021}}.
This work was supported by the Deutsche Forschungsgemeinschaft through the W\"urzburg-Dresden Cluster of Excellence on Complexity and Topology in Quantum Matter -- \textit{ct.qmat} (EXC 2147, Project No.\ 390858490). 
\end{acknowledgments}

\appendix

\section{Finite-size effects in the Luttinger-liquid phase\label{App:FSS_isolated}}

\begin{figure}[t]
    \centering
    \begin{overpic}[width=0.9\linewidth]{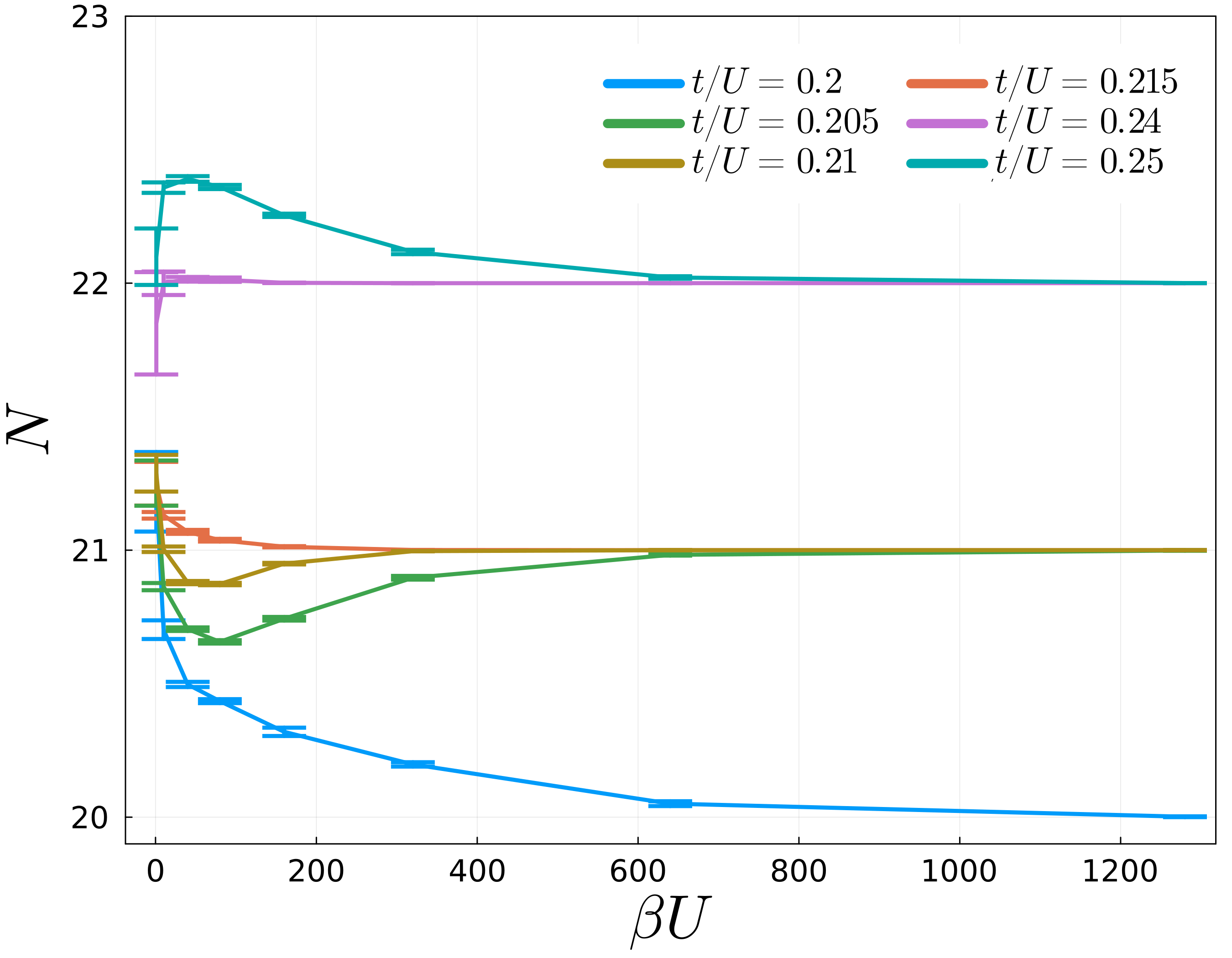}
    \put(15,67.5){\large(a)}
    \end{overpic}
    \begin{overpic}[width=0.9\linewidth]{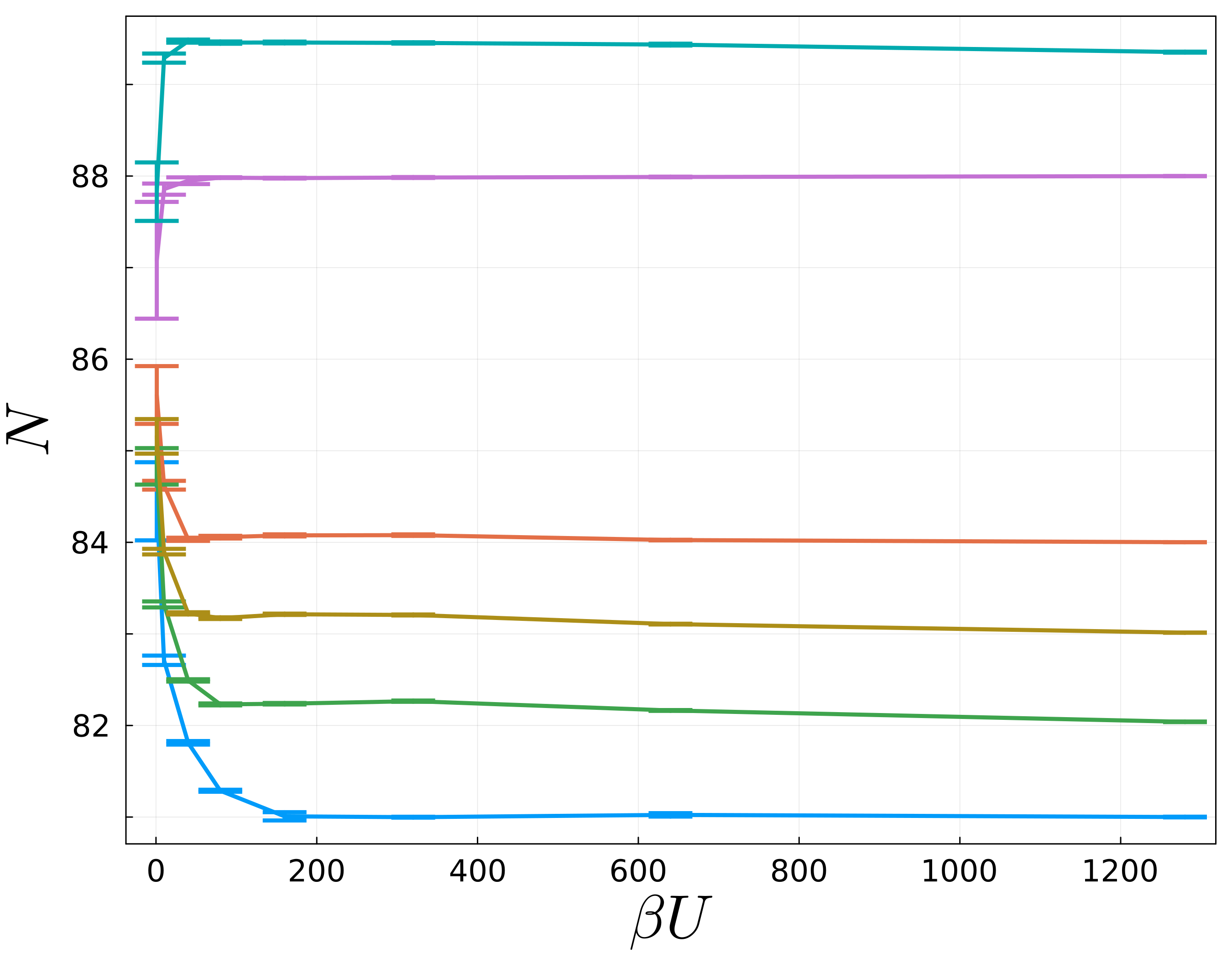}
    \put(15,67.5){\large(b)}
    \end{overpic}
    \caption{Total particle number of the closed system at fixed system sizes (a) $L=20$ and (b) $L=80$ as a function of inverse temperature $\beta$. Here, $\mu/U=0.3$.}
    \label{fig:particleconv}
\end{figure}

For all our QMC simulations, we make sure that results are converged to the ground state of our finite-size system. Within the Luttinger-liquid phase of the isolated system, finite-size effects can lead to an irregular dependence of observables on a tunable system parameter. We can easily understand these features by analyzing the temperature convergence of the total particle number $N$, as the hopping $t$ is changed at fixed $U$ and $\mu$. Figure \ref{fig:particleconv}(a) shows that for small system sizes of $L=20$ the particle number slowly converges to integer fillings as inverse temperature $\beta$ increases. In particular, for hopping parameters that are close by, \eg, $t/U\in\{0.205, 0.21, 0.215\}$, the system eventually converges to the same total particle number of $N=21$, although at intermediate finite temperatures $N$ can take different values.
By contrast, if we analyze $N(\beta)$ for the same parameters at a larger system size of $L=80$, as shown in Fig.~\ref{fig:particleconv}(b), every hopping parameter converges to a distinct integer filling (apart from $t/U=0.25$ which is still decaying very slowly). For a fixed distance between parameters, more possible fillings become accessible at larger $L$. Therefore, spurious finite-size effects as in Fig.~\ref{fig:particleconv}(a) would only reappear, if we increased the resolution in $t$. In addition, however, the possible absolute variations in $N/L$ decrease with increasing $L$, such that these effects become less severe at larger system sizes.
We observe that results for $N/L$ are essentially converged for $\beta U = 1280$. Finite-size effects originating from the particle-number sectors also affect other observables, but we find them to be less severe. In addition, estimators like the correlation length $\xi/L$ have significantly larger error bars than the particle number presented here, so that small variations will not play a role anymore.

If our system is coupled to a gapless dissipative bath the sort of finite-size effects discussed here get washed out quickly,  as we have seen in Fig.~\ref{fig:renorm}(a).  The smaller the bath exponent $s$, the quicker these effects seem to disappear. While in the isolated system, the ground state of our finite-size system has an integer particle number $N$, a small coupling to the bath lifts this quantization, as observed in Fig.~\ref{fig:renorm}(a). This behavior is related to an infinite compressibility that occurs for any coupling to the bath.

\begin{figure}[t]
    \centering
    \begin{overpic}[width=0.9\linewidth]{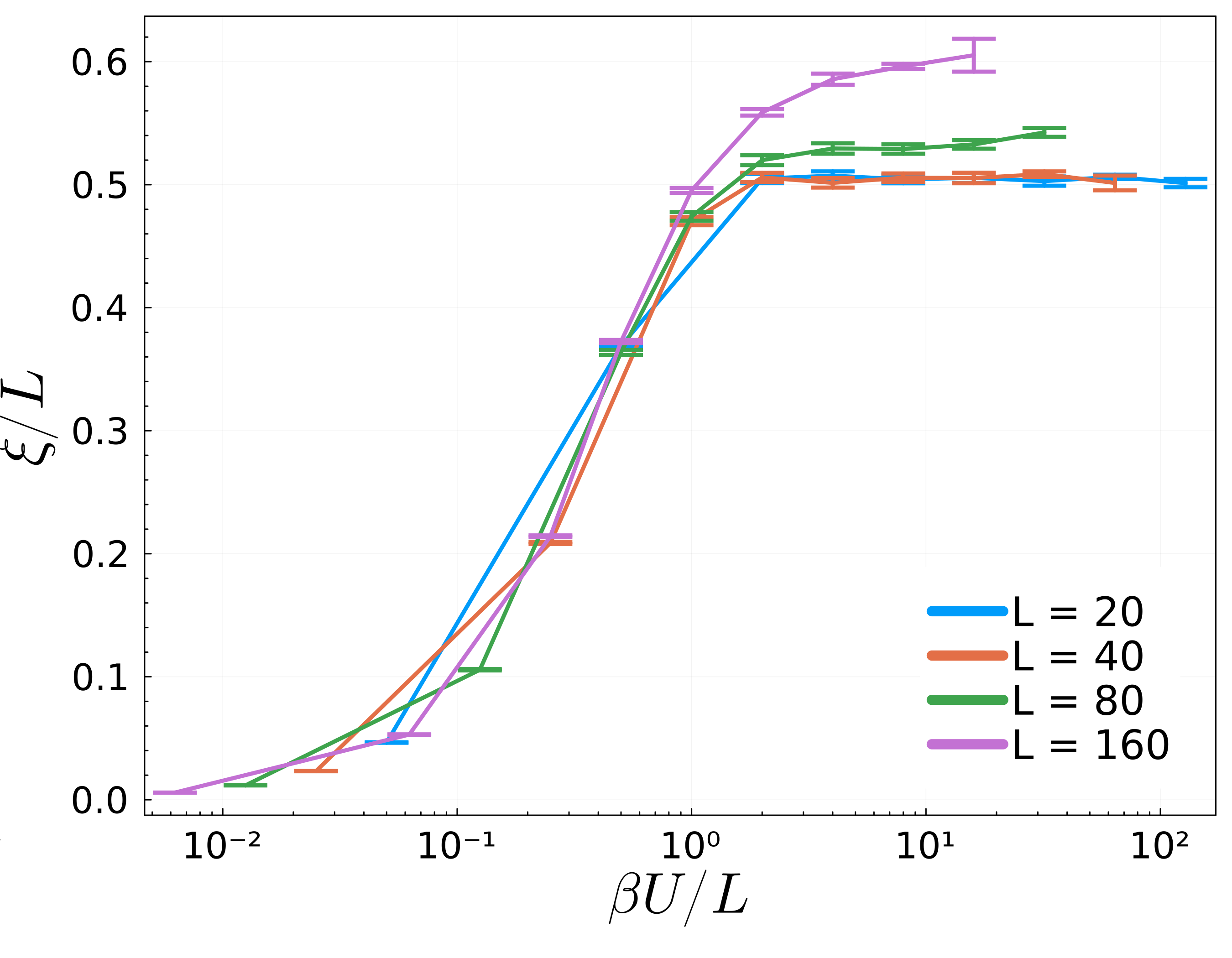}
    \put(15,67.5){\large(a)}
    \end{overpic}
    \begin{overpic}[width=0.9\linewidth]{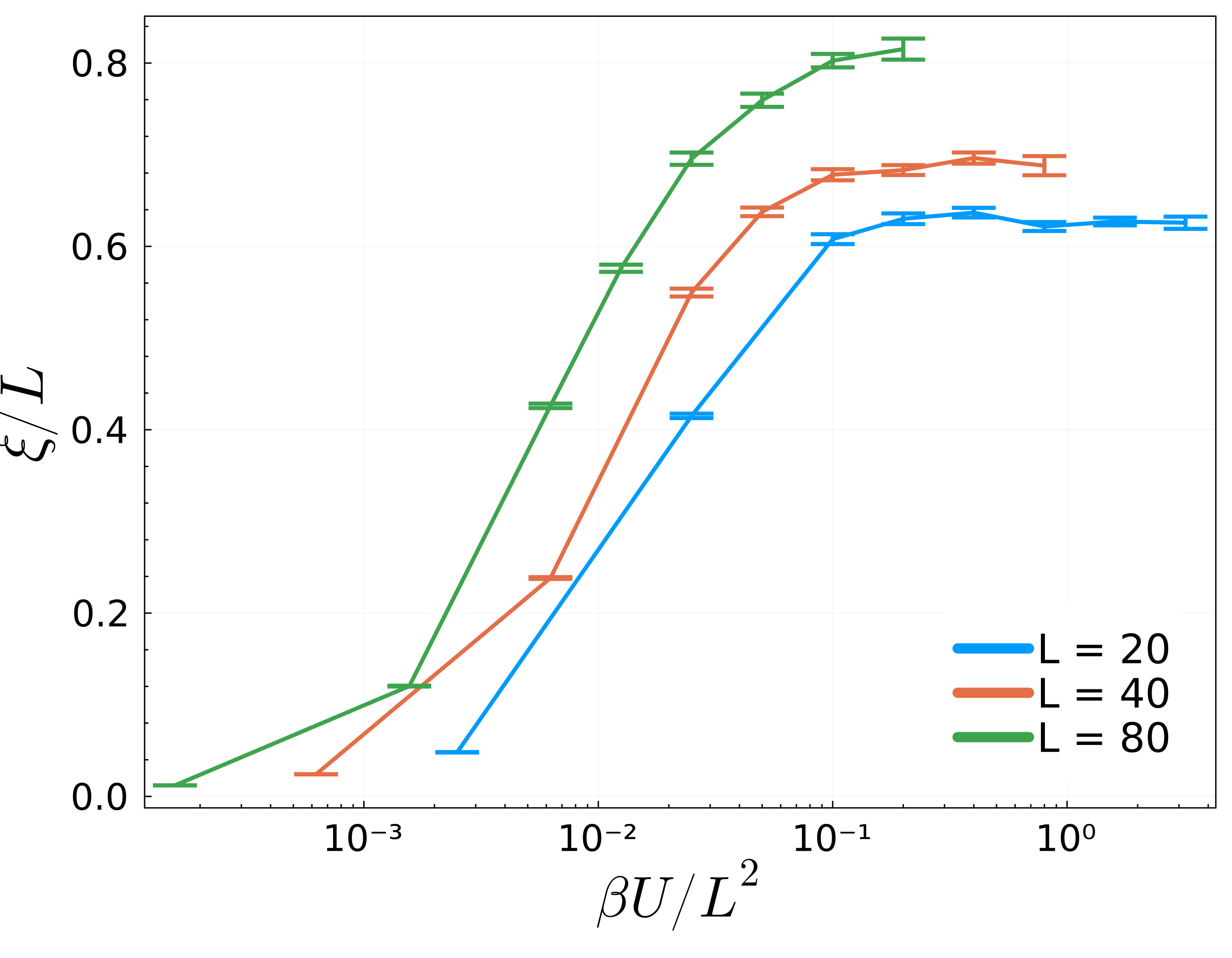}
    \put(15,67.5){\large(b)}
    \end{overpic}
    \caption{Temperature convergence starting from the Luttinger-liquid phase. We show the correlation length $\xi/L$ as a function of inverse temperature for dissipation strengths (a) $\alpha=0.02$ and (b) $\alpha=0.07$ and for different system sizes.
    Here, $t/U=0.25$ , $\mu/U=0.3$, and $s=1$. We performed simulations up to inverse temperatures of $\beta U =2560$ in panel (a) and $\beta U =1280$ in panel (b).}
    \label{fig:Tconvergence}
\end{figure}

\begin{figure*}[t]
    \centering
    \begin{overpic}[width=0.45\linewidth]{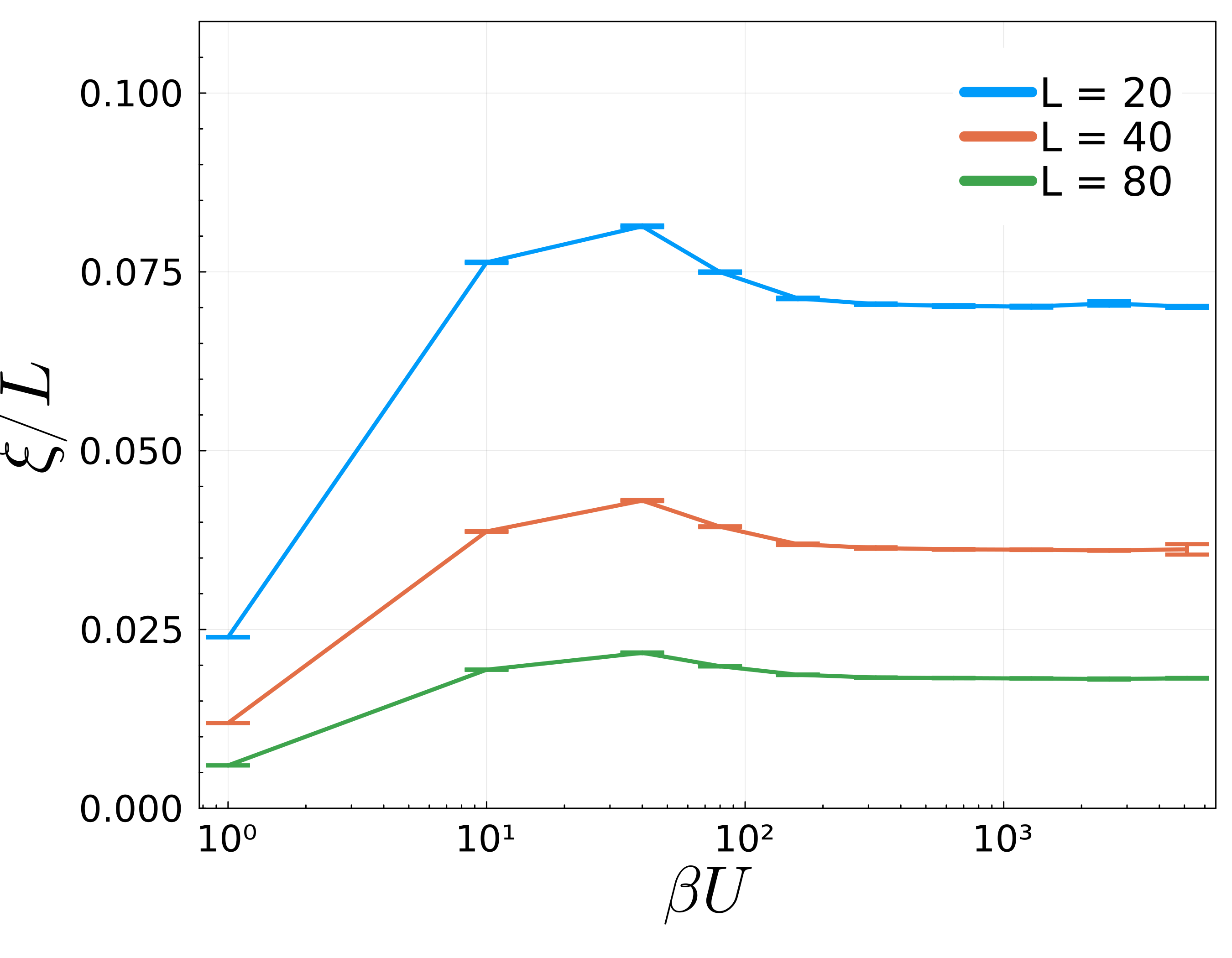}
    \put(88,16.5){\large(a)}
    \end{overpic}
    \hspace{0.5cm}
    \begin{overpic}[width=0.45\linewidth]{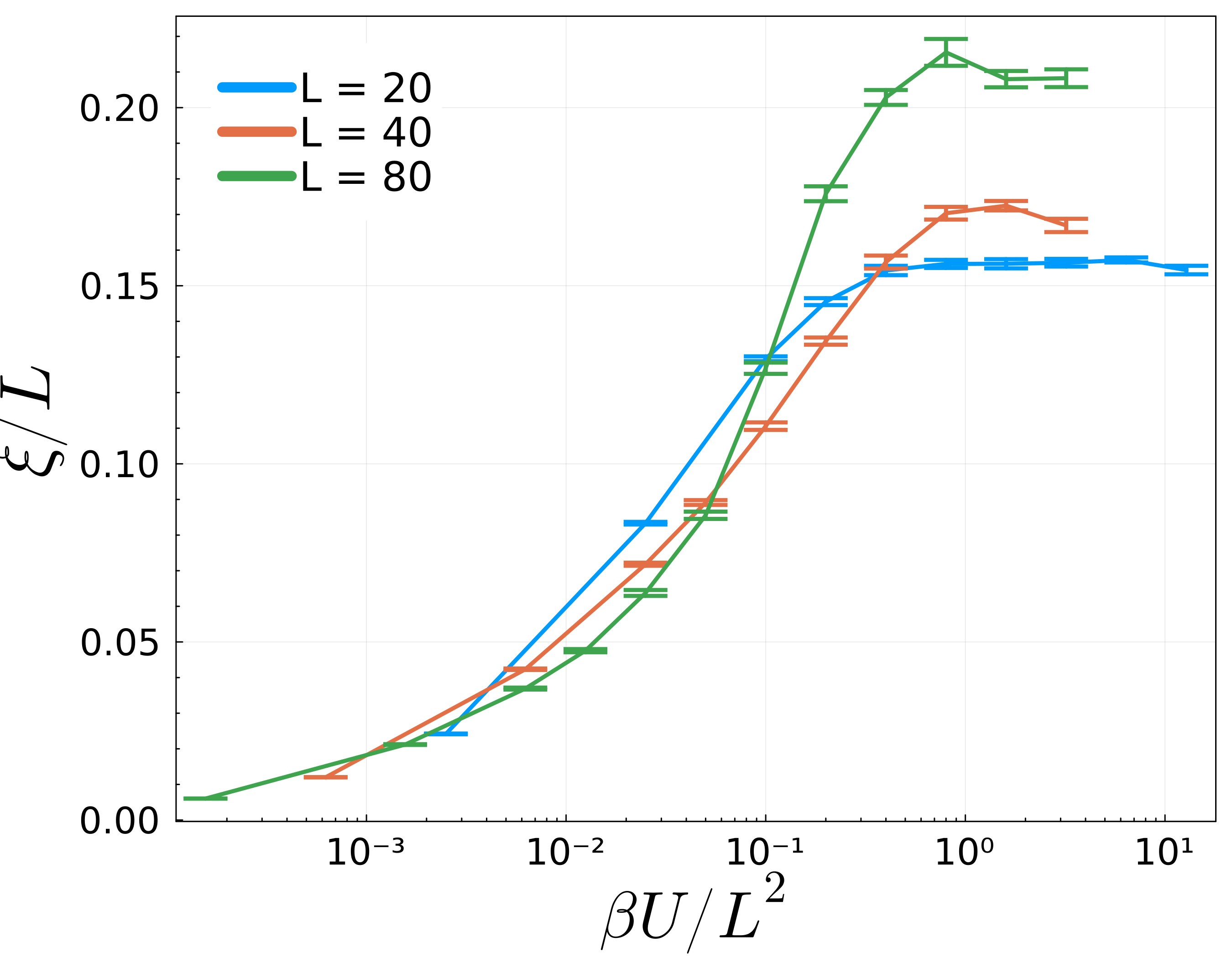}
    \put(88,15){\large(b)}
    \end{overpic}
    \caption{Temperature convergence starting from the Mott phase. We show the correlation length $\xi/L$ as a function of inverse temperature for dissipation strengths (a) $\alpha=0.04$ and (b) $\alpha=0.0625$ and for different system sizes.
    Here, $t/U=0.1$ , $\mu/U=0.3$, and $s=1$. We have performed simulations up to inverse temperatures of $\beta U = 5120$, only for $L=80$ in panel (b) we went up to $\beta U = 20480$.}
    \label{fig:TconvergenceMott}
\end{figure*}

\section{Temperature convergence for the Ohmic bath\label{App:convergence_ohmic}}

The long-range retarded interaction of the dissipative bath breaks conformal invariance in our model, so that it is not a priori clear which dynamical critical exponent $z$ to choose to perform the finite-size scaling. Therefore, we make sure that all our results are converged to the ground state of the finite-size system. Here, we analyze the temperature convergence of the correlation length $\xi / L$, which is our main observable to make quantitative predictions for the phases of the dissipative Bose-Hubbard model. We have convinced ourselves that the convergence of $\xi/L$ is representative for other observables.

Starting from the Luttinger-liquid phase of the isolated system, Fig.~\ref{fig:Tconvergence} shows $\xi/L$ as a function of inverse temperature $\beta$ and for two dissipation strengths. One can clearly see that with increasing system size we need to perform calculations at larger $\beta$ to obtain converged results. Eventually, the inverse temperatures above which our results reach a plateau follow $\beta \sim L^z$ where $z$ is determined by the corresponding phase. Our finite-size analysis of $\xi/L$ in Fig.~\ref{fig:corrcrossings} revealed that the system is in the long-range ordered phase for any $\alpha>0$, for which we expect to find $z=2$. However, the crossover towards this behavior might only become visible at large system sizes, in particular at weak coupling. Indeed, at $\alpha=0.02$ shown in Fig.~\ref{fig:Tconvergence}(a) our data is consistent with $z=1$ at small system sizes but the convergence shifts to larger $\beta$ for the biggest system sizes considered. By contrast, for a stronger coupling of $\alpha=0.07$ shown in Fig.~\ref{fig:Tconvergence}(b) the convergence in $\beta$ takes significantly longer than expected from $z=1$ and is consistent with $z=2$.

 We perform the same analysis again starting from the Mott phase. Deep in the insulating phase, the convergence in temperature is rather quick and does not show and system-size dependence, as can be seen from Fig.~\ref{fig:TconvergenceMott}(a). In the long-range ordered phase, but very close to the quantum phase transition, the convergence is much slower again and consistent with $z=2$, as far as one can conclude from only three system sizes [Fig.~\ref{fig:TconvergenceMott}(b)]. Even close to the transition, we have obtained converged results up to $L=80$. We have convinced ourselves that results converge even for larger $\alpha$.

\end{document}